%% file: main.tex
\documentclass[a4paper,11pt]{article}
\pdfoutput=1 

\usepackage{jheppub,bm,mathtools,slashed}
\usepackage[usenames,dvipsnames,svgnames,table,x11names]{xcolor}
\usepackage{extarrows}
\usepackage{easybmat}
\usepackage{nicematrix}
\usepackage{placeins}

\usepackage{hyperref} 
\usepackage{xurl}

\usepackage[T1]{fontenc} 
\usepackage[utf8]{inputenc}

\renewcommand*{\backref}[1]{}
\renewcommand*{\backrefalt}[4]{%
  \ifcase #1 %
    \relax 
  \or
    {\scriptsize (page~#2).}%
  \else
    {\scriptsize (pages~#2).}%
  \fi%
}

\usepackage[usenames,dvipsnames,svgnames]{xcolor}
\definecolor{refkey}{rgb}{0.9451,0.2706,0.4941}
\definecolor{labelkey}{rgb}{0.9451,0.2706,0.4941}

\definecolor{light_blue}{rgb}{0.15, 0.35, 0.95}
\definecolor{kit_green}{rgb}{0,
0.58823 
, 0.50980 
}

\usepackage{tikz}
\usepackage{pgfplots}
\pgfplotsset{compat=1.14}
\usepackage{tikz-3dplot}
\usetikzlibrary{intersections, calc,positioning,decorations.pathreplacing,decorations.pathmorphing,arrows,arrows.meta,patterns,pgfplots.fillbetween,math,matrix,shapes.geometric}
\tikzstyle{startstop} = [rectangle, rounded corners, minimum width=3cm, minimum height=1cm,text centered, draw=black, fill=red!30]
\tikzstyle{process} = [rectangle, minimum width=3cm, minimum height=1cm, text centered, draw=black, fill=blue!30]
\tikzstyle{decision} = [diamond, minimum width=3cm, minimum height=1cm, text centered, draw=black, fill=green!30]
\tikzstyle{arrow} = [thick,->,>=stealth]
\tikzstyle{box} = [rectangle, draw=black, minimum width=3cm, text centered]

\usepackage{multirow}
\usepackage{booktabs} 
\usepackage{siunitx} 
\usepackage{tabularx} 

\title{Data analysis of millions of Type IIB flux vacua}

\author[1,2,3]{Yuta Hamada,}
\author[4]{Yuma Uematsu,}
\author[3,5,6,7,8]{Masahito Yamazaki}

\affiliation[1]{Theory Center, IPNS, High Energy Accelerator Research Organization (KEK), 1-1 Oho, Tsukuba, Ibaraki 305-0801, Japan}
\affiliation[2]{Graduate Institute for Advanced Studies, SOKENDAI, 1-1 Oho, Tsukuba, Ibaraki 305-0801, Japan}
\affiliation[3]{RIKEN Center for Interdisciplinary Theoretical and Mathematical Sciences (iTHEMS), RIKEN, Wako 351-0198, Japan}
\affiliation[4]{High Energy Accelerator Research Organization (KEK), 1-1 Oho, Tsukuba, Ibaraki 305-0801, Japan}
\affiliation[5]{Graduate School of Science, University of Tokyo, Hongo 7-3-1, Tokyo 113-0033, Japan}
\affiliation[6]{Trans-Scale Quantum Science Institute, University of Tokyo, Hongo 7-3-1, Tokyo 113-0033, Japan}
\affiliation[7]{Kavli Institute for the Physics and Mathematics of the Universe (IPMU), UTIAS, University of Tokyo, Kashiwa 277-8583, Chiba 277-8583, Japan}
\affiliation[8]{Center for Data-Driven Discovery (CD3), Kavli IPMU, UTIAS, University of Tokyo, Kashiwa 277-8583, Chiba 277-8583, Japan}

\abstract{
We study the vacuum structure of flux compactifications of type IIB string theory on the degree 18 hypersurface in $\mathbb{CP}^4_{1,1,1,6,9}$ with two effective complex structure moduli. We scan $30{,}770{,}178$ flux choices compatible with the tadpole bound, and use a gradient-based constrained minimization based on the SLSQP algorithm to identify candidates for local minima of the scalar potential, yielding a dataset of millions of candidate flux vacua.
Using the $1{,}296{,}633$ reliable solutions selected from this dataset, we analyze distributions and correlations of various physical quantities such as vacuum expectation values of moduli and vacuum energy.
Using quadratic flux invariants as input features, we train a classifier to identify solutions satisfying the numerical SUSY criterion $|V|<10^{-4.5}$.
On a randomly selected test set, selecting the top-ranked $5\%$ recovers $47.2\%$ of the SUSY test solutions.
We also use supervised regression to predict the potential and moduli field values.
In the regression comparison, quadratic flux invariant inputs improve prediction of potential value and the complex-structure coordinates, while the vanilla six flux parameters give better predictions for the axio-dilaton coordinates under the tested settings.
}

\preprint{KEK-TH-2884, RIKEN-iTHEMS-Report-26}

\begin{document} 
\maketitle

\section{Introduction}\label{sec:intro}

It has been a long-standing dream of string theorists to connect the beauty of string theory to the low-energy physics describing our Universe.
One of the most crucial steps in this direction is to enumerate and analyze the vacua of string theory. There has been tremendous progress in recent years for higher-dimensional supersymmetric vacua, where evidence for string universality or string lamp-post principle has been accumulating~\cite{Adams:2010zy,Kim:2019vuc,Kim:2019ths,Cvetic:2020kuw,Montero:2020icj,Hamada:2021bbz,Bedroya:2021fbu,Cvetic:2022uuu,Font:2020rsk}.
For lower-dimensional vacua closer to our world, the number of vacua may be enormous~\cite{Douglas:2003um,Taylor:2015xtz}, but it is also conjectured to be finite~\cite{Acharya:2006zw,Hamada:2021yxy,Grimm:2023lrf,Delgado:2024skw}.
The finiteness implies that there might be hidden constraints or correlations among physical quantities in the low-energy effective theory obtained from string theory, in the spirit of the swampland program~\cite{Vafa:2005ui} (see e.g.~\cite{Palti:2019pca,vanBeest:2021lhn} for reviews).
To understand real-world physics from string theory, it is therefore important to figure out the properties and statistics of such vacua.
The statistical approach to type-IIB flux vacua was developed in \cite{Douglas:2003um,Ashok:2003gk,Denef:2004ze,Denef:2004cf}, where both counting formulae and distributions of supersymmetric and non-supersymmetric flux vacua were analyzed.

Among various approaches to construct four-dimensional string vacua, flux compactification is one of the promising directions~\cite{Klebanov:2000hb,Giddings:2001yu,Grimm:2004uq} given the rich structure of vacua and concrete proposals for moduli stabilization, such as the KKLT scenario~\cite{Kachru:2003aw,McAllister:2024lnt} and Large-Volume Scenario (LVS)~\cite{Balasubramanian:2005zx}.

The ensemble of flux vacua provides a huge dataset, which is an excellent playground for modern data science techniques.
Indeed, the numerical studies of flux vacua with modern techniques have recently been investigated~\cite{CaboBizet:2020cse,Cole:2021nnt,Dubey:2023dvu,Ebelt:2023clh,Krippendorf:2023idy,Chauhan:2025rdj,AbdusSalam:2025twp,Chauhan:2026gid,Chauhan:2026dss}. See also Refs.~\cite{Krippendorf:2025mhp,Walden:2025cpf} for applications of generative models (conditional variational autoencoders~\cite{NIPS2015} and Bayesian Flow Network~\cite{graves2025bayesianflownetworks}) for sampling flux vacua.\footnote{See Refs.~\cite{Halverson:2020opj,Seong:2024wkt,Yip:2025hon} for applications to string theory related topics.}

In this paper, we perform data analysis of flux vacua in type IIB Calabi-Yau orientifold compactifications with two effective complex structure moduli.
This is realized by a degree 18 hypersurface in the weighted projective space $\mathbb{CP}^4_{1,1,1,6,9}$.
We focus on a subsector with a discrete symmetry $\mathbb{Z}_6\times \mathbb{Z}_{18}$ so that the effective complex structure moduli space is two-dimensional.
We scan $30{,}770{,}178$ flux choices satisfying the tadpole bound and use a gradient-based constrained minimization based on the Sequential Least-Squares Quadratic Programming (SLSQP) algorithm.
This scan yields $46{,}388{,}212$ recorded endpoints of the numerical minimization, including both supersymmetric and non-supersymmetric candidates.
Applying the selection described in Section~\ref{sec:results} yields $1{,}296{,}633$ reliable solutions, including $127{,}241$ satisfying the numerical SUSY criterion $|V|<10^{-4.5}$.
While the SUSY solutions we have obtained are fewer than those in some previous studies (specifically designed to solve the F-term equations), the number of our \emph{non-SUSY solutions} is an order of magnitude larger than those in previous studies.
Note that while supersymmetric vacua are key ingredients for KKLT or LVS construction, non-supersymmetric vacua are also interesting in the context of de Sitter constructions~\cite{Saltman:2004sn,Hebecker:2020ejb,Carta:2021sms,Hebecker:2025tui} 
(which is a variant of dS uplift using small vacuum energy from complex structure and axio-dilaton sectors), and the various swampland conjectures regarding scalar potentials~\cite{Obied:2018sgi,Ooguri:2018wrx,Bedroya:2019snp}.

We also use supervised learning to classify SUSY and non-SUSY solutions and to predict the potential and moduli field values.
We compare models using the six flux parameters with models using quadratic flux invariants as inputs.

The paper is organized as follows.
In Section~\ref{sec:IIB_orientifold}, we review generalities on type IIB orientifold with three-form fluxes.
In Section~\ref{sec:model} we write down the explicit scalar potential for the concrete model with two complex structure moduli. 
Section~\ref{sec:results} describes the numerical search, sample construction and common numerical SUSY classification, followed by the distributions, correlations and flux dependence of the reliable solutions.
Section~\ref{sec:ml} studies classification and regression tasks on this sample.
We discuss the results and conclude in Section~\ref{sec:conclusion}.

\section{Type IIB orientifold}\label{sec:IIB_orientifold}
In this section, we review the type IIB orientifold of a Calabi-Yau threefold with O3/O7-planes and three-form fluxes.\footnote{For reviews of flux compactifications, see
\cite{Grana:2005jc,Douglas:2006es,Denef:2008wq}.}
We use the same notation as in \cite{Chauhan:2025rdj}.

The Hodge numbers of the Calabi-Yau threefold, $h^{1,1}$ and $h^{2,1}$, are split into even and odd parts under the orientifold action:
\begin{align}
    h^{1,1}=h^{1,1}_++h^{1,1}_- \;,
    \quad
    h^{2,1}=h^{2,1}_++h^{2,1}_- \;.
\end{align}
Correspondingly, there are $h_+^{1,1}$ K\"ahler moduli, and $h^{2,1}_-$ complex structure moduli that will be denoted by $Z^i$ ($i=1,\cdots,h^{2,1}_-$).

In this work, we focus on the complex structure moduli and the axio-dilaton, and do not address the stabilization or stability of the K\"ahler moduli.
For positive-energy local minima in this sector, a possible route to K\"ahler-moduli stabilization is suggested by Saltman and Silverstein~\cite{Saltman:2004sn}.
In this scenario, the positive flux contribution may be combined with nonperturbative and $\alpha'$ corrections to stabilize the K\"ahler moduli.

The classical K\"ahler potential $K$ for the complex structure moduli and the axio-dilaton is given by
\begin{align}
    K=-\log\left(-i\Pi^\dagger \,\Sigma \,\Pi\right) - \log\left(-i(\tau-\overline{\tau})\right) \;,
\label{eq:Kahler}\end{align}
where $\tau$ is the axio-dilaton, $\Sigma$ is a $(2h^{2,1}_-+2)\times (2h^{2,1}_-+2)$ matrix, and period vector $\Pi$ is a $(2h^{2,1}_-+2)$-dimensional vector:
\begin{align}
    \Sigma=
    \begin{pmatrix}
    \mathbf{0}_{h^{2,1}_-+1} & \mathbf{1}_{h^{2,1}_-+1} \\
    -\mathbf{1}_{h^{2,1}_-+1} & \mathbf{0}_{h^{2,1}_-+1}
    \end{pmatrix} \;,
    \quad \quad
    \Pi=
    \begin{pmatrix}
        2F-Z^iF_i\\
        F_i\\
        1\\
        Z^i
    \end{pmatrix} \;.
\label{eq:sigma_pi}\end{align}
Here $F$ is the prepotential (explicit expression will be shown in Eq.~\eqref{eq:prepotential}), and $F_i:=\partial_{Z^i}F$.

The Gukov-Vafa-Witten superpotential $W$ is~\cite{Gukov:1999ya}
\begin{align}
    W=\int G\wedge \Omega=-\left(f-\tau h\right)^T\,\Sigma\,\Pi \;.
\label{eq:GVW}\end{align}
Here $G=F_3-\tau H_3$ is the complexified three-form flux, and
\begin{align}
    &F_3 = m_I \alpha^I - m^I \beta_I \;,
    &&H_3 = n_I \alpha^I - n^I \beta_I \;,
    \label{eq:3form_flux}
\end{align}
are three-form fluxes expanded by the symplectic basis of three-form cohomology $\{\alpha^I,\beta_I\}$ ($I=0,\cdots,h^{2,1}_-$):
\begin{align}
    \int_M \alpha^I\wedge\beta_J=\delta^I_J \;,
    \quad
    \int_M \alpha^I\wedge\alpha^J=0 \;,
    \quad
    \int_M \beta_I\wedge\beta_J=0 \;.
\end{align}
The corresponding $(2h^{2,1}_-+2)$-dimensional flux vectors $f$ and $h$ are defined as
\begin{align}
    \label{eq:flux_general}
    f=(m_I,m^I) \;,\quad
    h=(n_I,n^I) \;,\quad
    m_I,m^I,n_I,n^I\in\mathbb{Z} \;.
\end{align}
Similarly, the holomorphic three-form $\Omega$ is expanded as
\begin{align}
    \Omega= F_I \alpha^I -  Z^I\beta_I \;,
\end{align}
where $F_I$ and $Z^I$ are defined as $\Pi=:(F_I,Z^I)^T$.

From \eqref{eq:Kahler} and \eqref{eq:GVW},\footnote{Due to the no-scale structure of K\"ahler moduli, $-3|W|^2$ term cancels out in the scalar potential. We suppress the overall K\"ahler-moduli-dependent factor.} the scalar potential is
\begin{align}
    V=e^K\left(K^{\tau\overline{\tau}}D_\tau W \overline{D_{\tau}W}+K^{i\bar j}D_iW\overline{D_{j}W}\right) \;,
    \label{eq:scalar_potential}
\end{align}
where $D_\tau W=\partial_\tau W+W\partial_\tau K$, $D_i W=\partial_{Z^i}W+W\partial_{Z^i}K$, and $K^{\tau\overline{\tau}}$, $K^{i\bar j}$ are the inverse of K\"ahler metric $K_{\tau\overline{\tau}}=\partial_\tau\partial_{\overline{\tau}}K$ and $K_{i\bar j}=\partial_{Z^i}\partial_{\overline{Z}^j}K$, respectively.
The string vacua correspond to local minima of $V$ as a function of $\tau$ and $Z^i$ for given values of fluxes $f$ and $h$.

Throughout this paper, we use the term supersymmetric (SUSY)
to refer to solutions satisfying the F-term conditions in the axio-dilaton and complex structure sector.
This terminology does not imply unbroken supersymmetry in the full theory including the K\"ahler sector.
The scalar potential \eqref{eq:scalar_potential} is positive semi-definite, and these F-term conditions are equivalent to $V=0$:
\begin{align}
    &D_\tau W
    =\frac{1}{\tau-\overline{\tau}}(f-\overline{\tau}h)^T\,\Sigma\,\Pi=0 \;,
    \nonumber\\
    &D_i W
    =-\left(f-\tau h\right)^T\,\Sigma\,\left(\partial_{Z^i}\Pi+\Pi\partial_{Z^i}K\right)=0 \;.
\label{eq:susy_condition}\end{align}
This condition is equivalent to imaginary self-dual (ISD) condition of three-form flux.
From the first equation, we observe
\begin{align}
    \overline{\tau}=-\frac{\left.W\right|_{\tau=0}}{\partial_\tau W} \;.
\end{align}
For supersymmetric solutions, we adopt the following normalization
convention for the vacuum value of the superpotential, based on Ref.~\cite{Demirtas:2019sip}:
\begin{align}
    W_0\coloneqq\sqrt{\frac{2}{\pi}}\left.e^{K/2}W\right|_{\text{vacuum}} \;.
    \label{eq:W0}
\end{align}
A small value $|W_0|\ll1$ is important for the KKLT construction~\cite{Kachru:2003aw}.

We have not yet specified the form of the prepotential $F$. In general, this requires solving Picard-Fuchs equations, but in the Large Complex Structure (LCS) limit, the prepotential takes the form
\begin{align}
    F=-\frac{\kappa_{ijk}}{6}Z^iZ^jZ^k+\frac{a_{ij}}{2}Z^iZ^j+b_iZ^i+\frac{i}{2}\tilde{\xi}+(\text{instanton corrections}) \;, \label{eq:prepotential}
\end{align}
Here $Y_3$ denotes the Calabi-Yau threefold whose complex-structure periods are described by $F$, and $\tilde{Y}_3$ denotes its mirror.
The constants $\kappa_{ijk}, a_{ij}$ and $b_i$ are real and determined by the topology of $\tilde{Y}_3$, and $\tilde{\xi}$ is
\begin{align}
    &\tilde{\xi}=\frac{\zeta(3)\chi(\tilde{Y}_3)}{(2\pi)^3} \;,
    &&\chi(\tilde{Y}_3)=2(\tilde{h}^{1,1}-\tilde{h}^{2,1}) \;.
\end{align}
The Hodge numbers $\tilde{h}^{1,1}$ and $\tilde{h}^{2,1}$ refer to $\tilde{Y}_3$.

The tadpole cancellation condition of D3-charge is
\begin{align}
    \frac{1}{2}N_\text{flux}+N_{D3}=\frac{Q_{D3}}{2} \;,
    \quad N_\text{flux} = \int H_3\wedge F_3= m_In^I - m^In_I \;,
    \label{eq:tadpole_cancellation}
\end{align}
where $N_{D3}$ is the number of D3-branes, and $Q_{D3}$ is the total D3-charge from orientifold projections.

\section{A model with two effective complex structure moduli}\label{sec:model}

We denote the smooth resolution of the degree-18 hypersurface in the weighted projective space $\mathbb{CP}^4_{1,1,1,6,9}$ by $X_3$; it has $(h^{1,1},h^{2,1})=(2,272)$.
We work in the $\Gamma=\mathbb{Z}_6\times\mathbb{Z}_{18}$ invariant sector.
In this sector, the effective complex-structure moduli are reduced to be two, which are denoted by $Z_1$ and $Z_2$.
This is the model we will numerically study in this paper.
The Greene-Plesser mirror~\cite{Greene:1990ud}, denoted by $Y_3$, is obtained by resolving $X_3/\Gamma$ and has Hodge numbers $(272,2)$.
The periods in the $\Gamma$-invariant sector of $X_3$ coincide with the two-parameter periods of $Y_3$~\cite{Candelas:1994hw,Demirtas:2019sip}.
Consequently, the topological data entering the effective prepotential are those of the mirror of $Y_3$, namely $\tilde{Y}_3=X_3$.
In particular, $\chi(X_3)=2(2-272)=-540$, giving the constant term $-i\,540\zeta(3)/[2(2\pi)^3]$.
The prepotential $F$ is~\cite{Candelas:1994hw,Martinez-Pedrera:2012teo}
\begin{align}
    F=-\frac{1}{6}(9Z_1^3+9Z_1^2Z_2+3Z_1Z_2^2)
    +\frac{1}{4}(9Z_1^2+6Z_1Z_2)
    +\frac{1}{4}(17Z_1+6Z_2)
    -\frac{i}{2}\frac{540\zeta(3)}{(2\pi)^3}
    +F_{\text{inst}} \;.
    \label{eq:prepotential_concrete}
\end{align}

We turn on the nonzero flux invariant under $\Gamma$, and the other fluxes are set to zero:
\begin{align}
    f = (m_A,0,\cdots,0,m^A,0,\cdots,0) \;, \quad
    h = (n_A,0,\cdots,0,0,n^A,0,\cdots,0) \;, \quad
    A=1,2,3 \;.
\end{align}
In the following, we omit writing zeros in the flux vectors for simplicity.
In this paper, we use an ansatz for the fluxes as\footnote{This is the same ansatz as in \cite{Martinez-Pedrera:2012teo} while this is not general enough~\cite{Chauhan:2025rdj}.}
\begin{align}
    &n^A = m_A \;, \quad
    n_A = - m^A \;,
    \nonumber\\
    &\text{equivalently, } \quad
    f = (m_A,-n_A) \;, \quad
    h = (n_A,m_A) \;.
    \label{eq:flux_ansatz}
\end{align}

The model has $SL(2,\mathbb{Z})$ duality symmetry acting as
\begin{align}
    &\tau\to\tau^\prime = \frac{a\tau+b}{c\tau+d} \;,
    \quad
    f\to f^\prime =
    \begin{pmatrix}
        b\,n_A + a\,m_A\\
        b\,m_A - a\,n_A
    \end{pmatrix} \;,
    \quad
    h\to h^\prime =
    \begin{pmatrix}
        d\,n_A+c\,m_A\\
        d\,m_A-c\,n_A
    \end{pmatrix} \;,
    \label{eq:SL(2,Z)}
\end{align}
where $ad-bc=1$, and $a,b,c,d \in \mathbb{Z}$.
In particular, we have the following transformations:
\begin{align}
    \begin{pmatrix}
        a & b \\ c & d
    \end{pmatrix} = 
    \begin{pmatrix}
        -1 & 0 \\ 0 & -1
    \end{pmatrix}:&
    \begin{pmatrix}
        m_A \\ n_A
    \end{pmatrix}\mapsto
    \begin{pmatrix}
        -m_A \\ -n_A
    \end{pmatrix} \;,
    \quad \tau\mapsto \tau \;,
    \\
    \begin{pmatrix}
        a & b \\ c & d
    \end{pmatrix} = 
    \begin{pmatrix}0 & -1 \\ 1 & 0
    \end{pmatrix}:&
    \begin{pmatrix}
        m_A \\ n_A
    \end{pmatrix}\mapsto
    \begin{pmatrix}
        -n_A \\ m_A
    \end{pmatrix} \;,
    \quad \tau\mapsto -\frac{1}{\tau} \;,
    \\
    \begin{pmatrix}
        a & b \\ c & d
    \end{pmatrix} = 
    \begin{pmatrix}0 & 1 \\ -1 & 0
    \end{pmatrix}:&
    \begin{pmatrix}
        m_A \\ n_A
    \end{pmatrix}\mapsto
    \begin{pmatrix}
        n_A \\ -m_A
    \end{pmatrix} \;,
    \quad \tau\mapsto -\frac{1}{\tau} \;.
\end{align}
Namely, the transformations generated by $-I$ and $S$ do not change the flux ansatz.
This indicates that we can take $m_1\geq0$ and $n_1\geq0$ without loss of generality.

The tadpole cancellation condition~\eqref{eq:tadpole_cancellation} bounds the fluxes as:
\begin{align}
    m_1^2+m_2^2+m_3^2+n_1^2+n_2^2+n_3^2\leq Q_{D3} \;.
\end{align}
Note that $Q_{D3}=276$ if we use the orientifold constructed in \cite{Louis:2012nb,Dubey:2023dvu}.

In this model, the superpotential, its derivatives, and K\"ahler derivatives are explicitly written as
\input{W.tex}
and 
\input{dK.tex}

The general solution is obtained by (locally) minimizing the scalar potential~\eqref{eq:scalar_potential}
\begin{align}
    V=e^K\left(K^{\tau\overline{\tau}}D_\tau W \overline{D_{\tau}W}+\sum_{i,j=Z_{1,2}}K^{i\bar j}D_iW \overline{D_{j}W}\right) \;,
\end{align}
where
\input{K.tex}
The instanton correction to the prepotential~\eqref{eq:prepotential_concrete} is given by
\begin{align}
    &F_\text{inst} = \frac{1}{(2\pi i)^3} \bigg( -540q_1 - 3 q_2 - \frac{1215}{2}q_1^2 + 1080q_1 q_2 +\frac{45}{8}q_2^2 +\cdots\bigg) \;,
    &&q_a = e^{2\pi i Z_a} \;,
    \label{eq:F_inst}
\end{align}
where $a=1,2$.

This two-parameter model has been used in several previous numerical studies of flux vacua. We will compare our scan with these works in Section~\ref{sec:results}.
For earlier explicit constructions and statistical analyses of flux vacua in related Calabi-Yau orientifolds, see Refs.~\cite{Giryavets:2003vd,Conlon:2004ds,Blanco-Pillado:2020hbw} for example.

\section{Numerical Results}\label{sec:results}

In this section, we perform a numerical search for flux vacua in the model of Section~\ref{sec:model} and analyze the resulting dataset.
Section~\ref{sec:ml} uses the same sample for supervised prediction.

\subsection{Numerical search and dataset construction}\label{subsec:ml_data}

\begin{figure}[t]
\centering
\begin{tikzpicture}[node distance=1.1cm]
    \node (flux) [box, align=center, text width=0.78\linewidth]
    {Choose a flux vector
    $\vec{N}=(m_1,m_2,m_3,n_1,n_2,n_3)$\\
    satisfying the tadpole bound \eqref{eq:tadpole_cancellation}};

    \node (potential) [box, below of=flux, yshift=-0.5cm, align=center, text width=0.78\linewidth]
    {Construct $K$, $W$, and the no-scale potential $V$\\
    from \eqref{eq:Kahler}, \eqref{eq:GVW}, and \eqref{eq:prepotential_concrete}};

    \node (slsqp) [box, below of=potential, yshift=-0.5cm, align=center, text width=0.78\linewidth]
    {Run SLSQP minimization from 10 random initial points\\
    for \(\vec{x}=(\mathrm{Re}\,\tau,\mathrm{Im}\,\tau,
    \mathrm{Re}\,Z_1,\mathrm{Im}\,Z_1,
    \mathrm{Re}\,Z_2,\mathrm{Im}\,Z_2)\)};

    \node (check) [box, below of=slsqp, yshift=-0.5cm, align=center, text width=0.78\linewidth]
    {Require successful convergence, 
    $\max_A|\partial V/\partial x_A|\le 0.1$};

    \node (post) [box, below of=check, yshift=-0.5cm, align=center, text width=0.78\linewidth]
    {Apply the preliminary duplicate check within each flux choice\\
    by rounding the potential and field values to two significant digits};

    \node (reliable) [box, below of=post, yshift=-0.5cm, align=center, text width=0.78\linewidth]
    {Impose $\mathrm{Im}\,Z_i\geq 1$,\\
    and Hessian condition $\lambda_{\min}(H)\geq-\delta$\\
    using the tolerance in \eqref{eq:hessian_cut}};

    \node (dedup) [box, below of=reliable, yshift=-0.65cm, align=center, text width=0.78\linewidth]
    {Remove numerical duplicates within each $\mathbb{Z}_4$ flux orbit\\
    using the K\"ahler-metric criterion $D_K\leq10^{-3}$, Eq.~\eqref{eq:dedup_distance}};

    \node (classify) [box, below of=dedup, yshift=-0.5cm, align=center, text width=0.75\linewidth]
    {Classify reliable solutions by the criterion\\
    $|V|<V_*$ SUSY, \quad $|V|\geq V_*$ non-SUSY, Eq.~\eqref{eq:vacuum_criterion}};

    \draw [arrow] (flux) -- (potential);
    \draw [arrow] (potential) -- (slsqp);
    \draw [arrow] (slsqp) -- (check);
    \draw [arrow] (check) -- (post);
    \draw [arrow] (post) -- (reliable);
    \draw [arrow] (reliable) -- (dedup);
    \draw [arrow] (dedup) -- (classify);
\end{tikzpicture}
\caption{
Numerical pipeline for the flux scan. The Hessian cut also retains solutions with $|\lambda_{\min}(H)|\leq\delta$.
}
\label{fig:flowchart}
\end{figure}
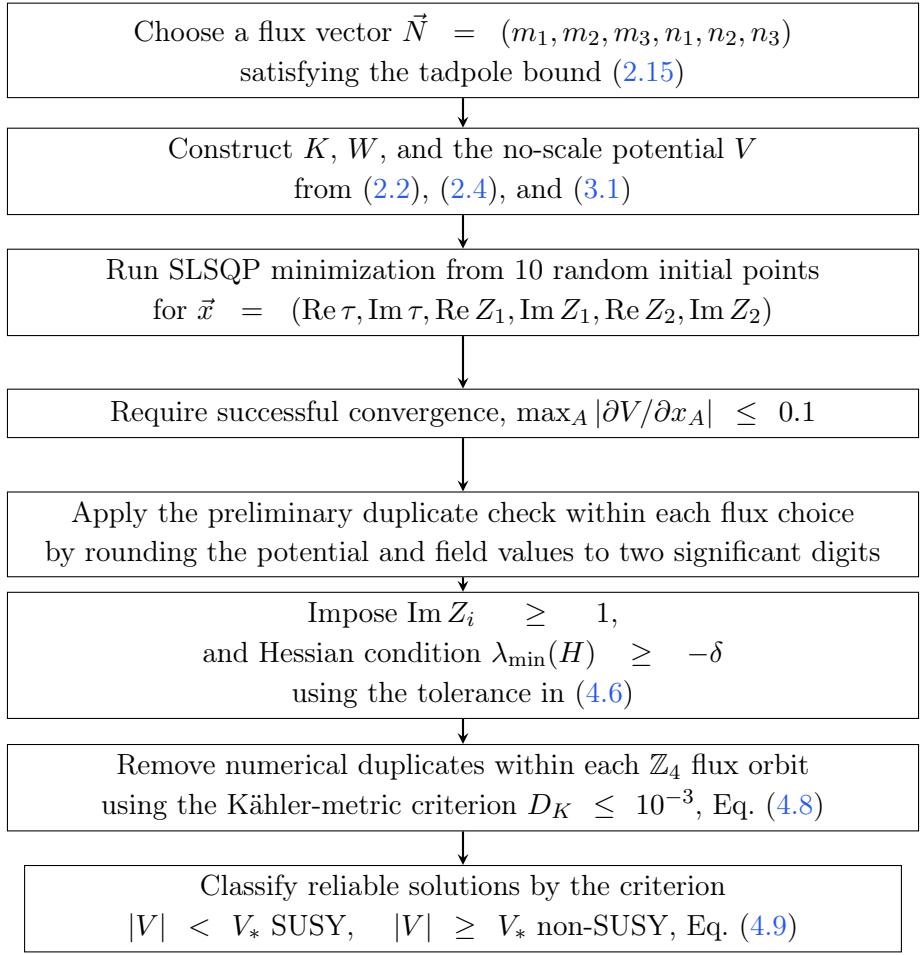

The overall pipeline of our scan is given in Figure~\ref{fig:flowchart}.\footnote{The scripts for the numerical computation are available in Ref.~\cite{FluxVacuaSearch}.}
The range of fluxes we have studied is
\begin{align}
    m_1, n_1&=0,\cdots,16 \;,
    && m_2, n_2= -16,\cdots,16 \;,
    && m_3, n_3=-16,\cdots,16 \;,
\end{align}
with the tadpole cancellation condition
\begin{align}
    m_1^2+m_2^2+m_3^2+n_1^2+n_2^2+n_3^2\leq 276 \;.
\end{align}
There are $30{,}770{,}178$ combinations of fluxes in total.
For each choice of fluxes, we minimize the scalar potential \eqref{eq:scalar_potential} as a function of the six real coordinates of $\tau,Z_1,Z_2$.
The minimization is performed by the SLSQP algorithm implemented in \texttt{scipy.\allowbreak optimize.\allowbreak minimize}~\cite{Virtanen:2019joe}.
We use the scalar potential as the objective function and provide its gradient to the optimizer.
The SLSQP stopping tolerance \texttt{ftol} is set to $10^{-5}$ via the \texttt{tol} argument of \texttt{scipy.\allowbreak optimize.\allowbreak minimize}.
Since the scalar potential can have multiple local minima for a fixed choice of fluxes, we repeat the minimization ten times with different random initial conditions.
In each trial, the initial values are drawn independently from
\begin{align}
    &\mathrm{Re}\,\tau,\mathrm{Re}\,Z_1,\mathrm{Re}\,Z_2 \in [-2,2] \;,
    &&\mathrm{Im}\,\tau,\mathrm{Im}\,Z_1,\mathrm{Im}\,Z_2 \in [0.1,2] \;.
\end{align}
The chosen initial-value range and the finite number of trials may bias which solutions are found, and the failure to find a solution for a given flux choice does not establish the absence of such solutions.

During the minimization, we impose the positivity constraints
\begin{align}
    &\mathrm{Im}\,\tau\geq 10^{-5} \;,\qquad
    &&\mathrm{Im}\,Z_1\geq 10^{-5} \;,\qquad
    &&\mathrm{Im}\,Z_2\geq 10^{-5} \;.
\end{align}
In order to select solutions for analysis, we require successful convergence, and a sufficiently small gradient. More explicitly, for a solution $\vec{x}_*$, we impose
\begin{align}
    &\max_A\left|
    \frac{\partial V}{\partial x_A}(\vec{x}_*)
    \right| \leq 0.1 \;,
    &&\vec{x} = (\mathrm{Re}\,\tau,\mathrm{Im}\,\tau,
   \mathrm{Re}\,Z_1,\mathrm{Im}\,Z_1,
   \mathrm{Re}\,Z_2,\mathrm{Im}\,Z_2) \;.
\label{eq:gradient_selection}
\end{align}

Different random initial conditions may converge to the same solution. During the initial scan, we apply a preliminary duplicate check within each flux choice: two recorded endpoints are identified when their potential and field values agree after rounding each component to two significant digits.
We apply the additional duplicate removal described below after the selection cuts.
Although the axio-dilaton can be mapped to the $SL(2,\mathbb{Z})$ fundamental domain using~\eqref{eq:SL(2,Z)}, the statistics and figures in this and next sections use the saved coordinates before this mapping.
To make this full scan computationally feasible, the batch-job configuration was tuned to account for queueing overhead and process-count constraints as well as raw CPU performance.
Using this setup, we scanned all combinations within a wall-clock time of order one week and found $46{,}388{,}212$ recorded endpoints.

In the numerical scan, we use the polynomial large complex structure prepotential~\eqref{eq:prepotential}, including the perturbative $\zeta(3)$ correction, but neglect $F_{\rm inst}$~\eqref{eq:F_inst}.
To suppress instanton corrections, we require $\mathrm{Im}\,Z_i\geq1$ in addition to the selection criteria above, leaving $2{,}309{,}556$ solutions.
For these solutions, we evaluate the Hessian $H_{AB}=\partial_A\partial_B V$ in the six real variables $\vec{x}$ and exclude solutions with
\begin{align}
    \lambda_{\min}(H)<-\delta,\qquad
    \delta=10^{-10}+10^{-5}\max_i|\lambda_i(H)| \;.
    \label{eq:hessian_cut}
\end{align}
This removes $14{,}220$ solutions, leaving $2{,}295{,}336$ solutions.
We next remove numerical duplicates, including those related by the $\mathbb{Z}_4$ action $(\mathbf m,\mathbf n)\mapsto(-\mathbf n,\mathbf m)$ and $\tau\mapsto-1/\tau$, with $\mathbf m=(m_1,m_2,m_3)$, $\mathbf n=(n_1,n_2,n_3)$.
Choosing a representative of each $\mathbb{Z}_4$ orbit, we compare the field values using the K\"ahler metric.
In units $M_{\rm Pl}=1$, we use
\begin{align}
    ds^2=2K_{\tau\bar\tau}\,d\tau\,d\bar\tau
    +2\sum_{i,j=1}^{2}K_{i\bar j}\,dZ_i\,d\bar Z_j.
    \label{eq:dedup_metric}
\end{align}
For two field configurations $a,b$, by setting $\Delta Z_i=Z_{i,b}-Z_{i,a}$ and $Z_i(t)=Z_{i,a}+t\Delta Z_i$, we define
\begin{align}
    d_\tau(a,b)&=\sqrt{2}\operatorname{arsinh}
    \frac{|\tau_a-\tau_b|}{2\sqrt{\mathrm{Im}\,\tau_a\,\mathrm{Im}\,\tau_b}},
    \nonumber\\
    L_{\rm CS}(a,b)&=\int_0^1 dt\,
    \left[2\sum_{i,j=1}^{2}K_{i\bar j}(Z(t))
    \Delta Z_i\Delta\bar Z_j\right]^{1/2},
    \nonumber\\
    D_K(a,b)&=\sqrt{d_\tau(a,b)^2+L_{\rm CS}(a,b)^2}.
    \label{eq:dedup_distance}
\end{align}
Here $d_\tau$ is the exact axio-dilaton geodesic distance.
We approximate the complex-structure geodesic distance $d_{\rm CS}$ by $L_{\rm CS}$, where the path is taken to be the straight line.
Since this path need not be the shortest, we have $d_{\rm CS}\leq L_{\rm CS}$.
Therefore the exact full geodesic distance $d_{\rm geo}$ satisfies $d_{\rm geo}=\sqrt{d_\tau^2+d_{\rm CS}^2}\leq D_K$.

Within each $\mathbb{Z}_4$ flux orbit, we group solutions so that every pair in a group satisfies $D_K\leq10^{-3}$ and keep the single solution in each group.
This removes $998{,}703$ duplicate records and leaves $1{,}296{,}633$ reliable solutions. 
All statistics and data-based figures in this and the next sections use this sample.
For each solution, we record the six flux parameters $\vec N$, the potential $V$, and the six coordinates $\vec x_*$.
There are $1{,}167{,}799$ distinct flux vectors and $1{,}147{,}948$ distinct $\mathbb Z_4$ flux orbits. 

We also checked for additional identifications under the LCS monodromies $Z_i\to Z_i+k_i$, with $(k_1,k_2)\in\mathbb{Z}^2\setminus\{(0,0)\}$, including the corresponding transformations of both flux vectors.
No flux configurations in our dataset are related by such shifts, even when combined with arbitrary $SL(2,\mathbb{Z})$ transformations.

\begin{figure}[htbp]
    \centering
    \includegraphics[width=0.85\textwidth]{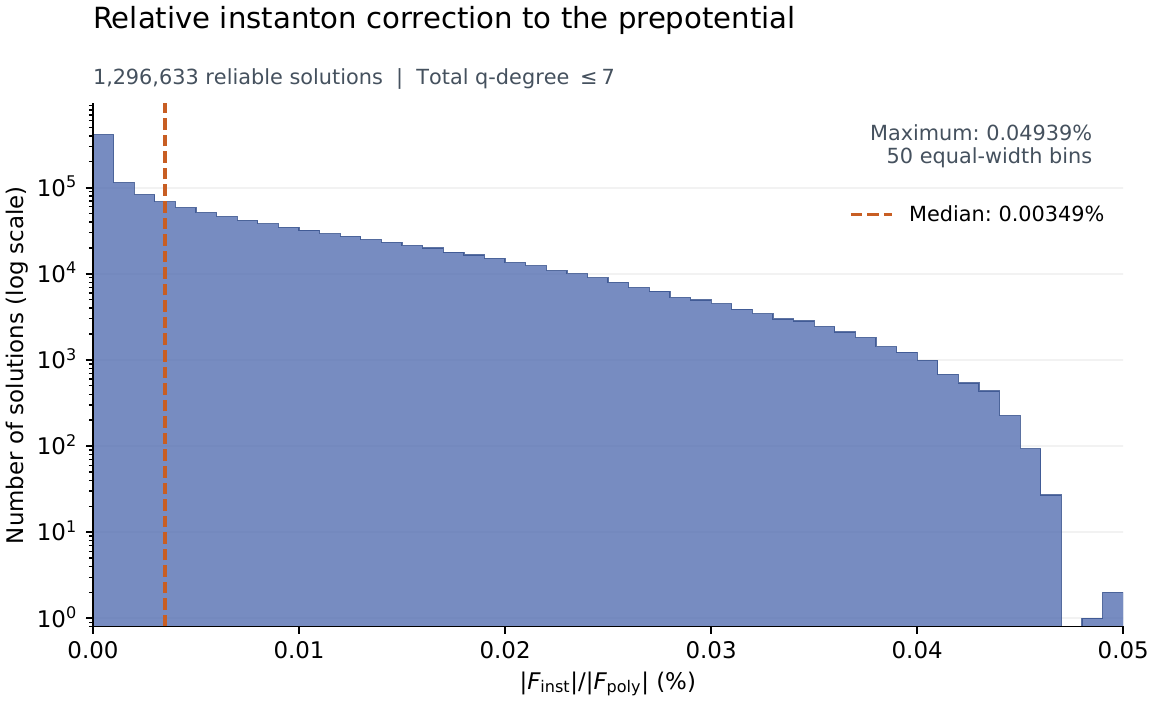}
    \caption{Distribution of $|F_{\rm inst}^{(\leq7)}|/|F_{\rm poly}|$ for reliable solutions. 
    The horizontal axis is the ratio, and the vertical axis shows the number of solutions on a logarithmic scale. 
    The dashed line corresponds to the median.}
    \label{fig:F_correction_histogram}
\end{figure}

To quantify the size of the instanton correction, we evaluate the ratio $|F_{\rm inst}^{(\leq7)}|/|F_{\rm poly}|$ at the field values of $1{,}296{,}633$ reliable solutions.
Here $F_{\rm poly}$ is the polynomial part of \eqref{eq:prepotential_concrete}, including the $\zeta(3)$ term, and $F_{\rm inst}^{(\leq7)}$ includes all terms $q_1^a q_2^b$ with $1\leq a+b\leq7$, using the instanton numbers in Ref.~\cite[Table~7.1]{Candelas:1994hw}.
Figure~\ref{fig:F_correction_histogram} shows the distribution of this ratio.
The median and maximum are $0.00349\%$ and $0.0494\%$, respectively.
Thus, the instanton correction to the prepotential is small throughout the dataset.

In Table~\ref{tab:stats_reliable_fd_I}, we summarize the statistics of the reliable solutions.
In the following, we study the detailed distributions of the potential and moduli, their correlations, and the flux dependence and multiplicity of the solutions.

\begin{table}[h]
  \centering
  \scriptsize
  \setlength{\tabcolsep}{5.5pt}
  \begin{tabular}{lrrrrrrr}
    \toprule
    &
    $V$ &
    $\mathrm{Re}\,\tau$ & $\mathrm{Im}\,\tau$ &
    $\mathrm{Re}\,Z_1$ & $\mathrm{Im}\,Z_1$ &
    $\mathrm{Re}\,Z_2$ & $\mathrm{Im}\,Z_2$ \\
    \midrule
    min &
    $2.25\times10^{-10}$ & $-3.52\times10^{13}$ & $1.08\times10^{-3}$ & $-2.60\times10^{11}$ & 1.00 & $-3.37\times10^{13}$ & 1.00 \\
    25\% &
    $1.84\times10^{1}$ & -1.06 & 0.306 & -0.507 & 1.10 & -1.40 & 1.38 \\
    50\% &
    $6.53\times10^{1}$ & -0.428 & 0.650 & 0.287 & 1.24 & 0.642 & 2.12 \\
    75\% &
    $1.14\times10^{2}$ & 0.129 & 1.39 & 0.984 & 1.52 & 2.41 & 3.04 \\
    max &
    $1.27\times10^{7}$ & $1.42\times10^{12}$ & $2.29\times10^{13}$ & $1.18\times10^{13}$ & $2.17\times10^{12}$ & $4.52\times10^{11}$ & $9.17\times10^{13}$ \\
    \bottomrule
  \end{tabular}
  \caption{Summary statistics for the $1{,}296{,}633$ reliable solutions as described in Section~\ref{subsec:ml_data}.
  }
  \label{tab:stats_reliable_fd_I}
\end{table}

\subsection{Potential distribution and numerical SUSY classification}\label{subsec:ml_criterion}

\begin{figure}[ht]
 \centering
 \includegraphics[width=0.92\textwidth]{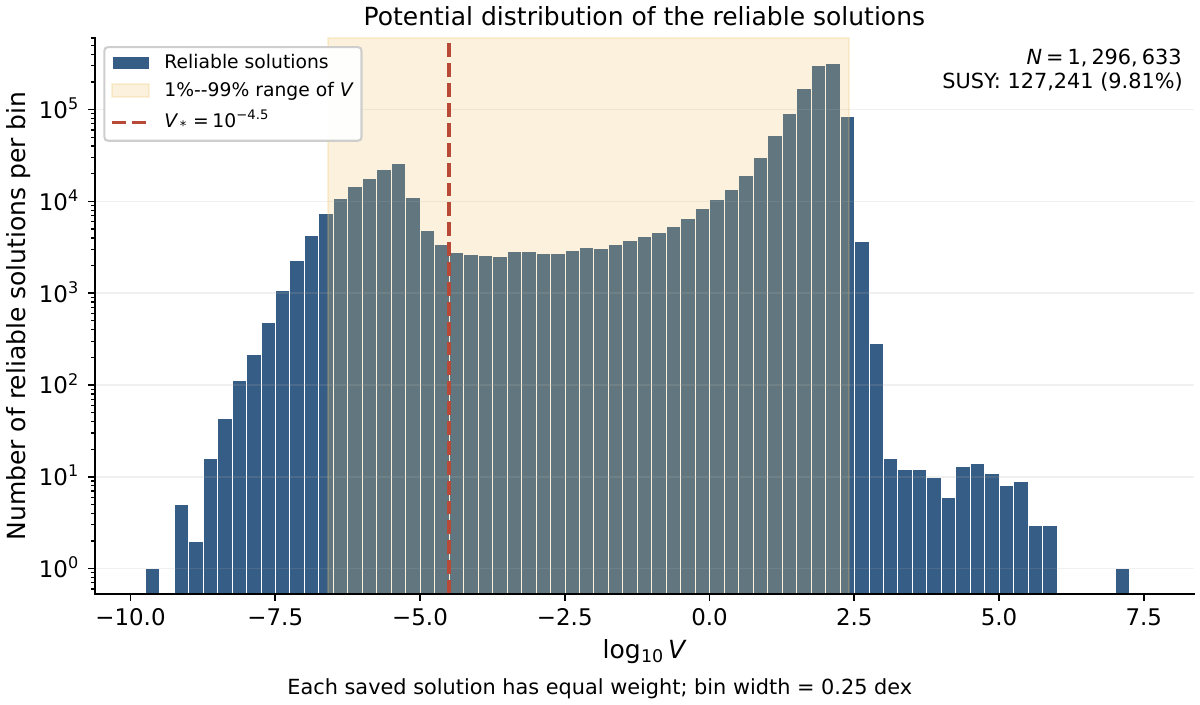}
 \caption{Histogram of $\log_{10}V$ reliable solutions, with bin width $0.25$. 
 The dashed line is the numerical SUSY criterion $V_*=10^{-4.5}$.
 The shaded range lies between the $1\%$ and $99\%$ quantiles.}
 \label{fig:ml_hist_logV}
\end{figure}

Figure~\ref{fig:ml_hist_logV} shows the potential distribution on a logarithmic scale with two peaks separated by a lower density region.
From this figure, we adopt the numerical SUSY criterion
\begin{align}
 |V|<V_*\equiv10^{-4.5}\simeq3.16\times10^{-5},
 \label{eq:vacuum_criterion}
\end{align}
throughout this work.
This criterion selects $127{,}241$ SUSY solutions ($9.81\%$ of the sample), and $1{,}169{,}392$ non-SUSY solutions.

In Table~\ref{tab:vacua_counts}, we compare the reported solution counts and search conditions for the two-modulus model in earlier studies.
\begin{table}[p]
  \centering
  \footnotesize
  \setlength{\tabcolsep}{3pt}
  \renewcommand{\arraystretch}{1.08}
  \begin{tabularx}{\textwidth}{@{}>{\raggedright\arraybackslash}p{0.20\textwidth}>{\raggedright\arraybackslash}p{0.23\textwidth}>{\raggedright\arraybackslash}X>{\raggedright\arraybackslash}p{0.22\textwidth}@{}}
    \toprule
    Study and count & Tadpole and flux choices & Selection and approximation & Duplicate identification \\
    \midrule
    \cite[Sec.~3.3]{Martinez-Pedrera:2012teo}\newline
    F-term roots: $500{,}865$\newline
    After cuts: $24{,}882$
    & $N_{\rm flux}\leq34$; ansatz~\eqref{eq:flux_ansatz}.
    & F-term equations; polynomial prepotential; $g_s>0$; LCS test (tolerance $0.1$). 
    & $\mathbb{Z}_4$ identification; $\tau$ mapped to its fundamental domain.
    \\
    \addlinespace[4pt]
    \cite[Sec.~4.1]{Dubey:2023dvu}\newline
    SUSY: $33{,}019$ (polynomial), $34{,}542$ (with instantons)
    & No cut on $N_{\rm flux}$; twelve-component fluxes.
    & F-term equations; polynomial prepotential with or without instanton corrections through degree 10; $g_s>0$; mirror K\"ahler-cone check; Hessian check; LCS test (tolerance $0.1$).
    & $SL(2,\mathbb Z)$ identification.
    \\
    \addlinespace[4pt]
    \cite[Sec.~3.1]{Krippendorf:2023idy}\newline
    Non-SUSY: $77{,}849$
    & $N_{\rm flux}\leq276$; uniform integer-flux sampling. 
    & Critical points; polynomial prepotential with instanton corrections through degree 10; Hessian check. 
    & $SL(2,\mathbb Z)$ identification in the referenced framework.
    \\
    \addlinespace[4pt]
    \cite[Table~1]{Chauhan:2025rdj}\newline
    SUSY:\newline
    A: $5{,}140{,}872$\newline
    B: $12{,}196$\newline
    C: $50{,}884{,}086$\newline
    D: $123{,}408{,}240$
    & $N_{\rm flux}\leq N_{\max}$, with $N_{\max}=34,10,34,50$ for A--D. Twelve-component fluxes.
    & F-term equations; polynomial prepotential; LCS tests: ratios $\leq10^{-5}$.
    & $SL(2,\mathbb Z)$ and complex-structure monodromy identification.
    \\
    \addlinespace[4pt]
    This work\newline
    Non-SUSY: $1{,}169{,}392$\newline
    SUSY: $127{,}241$
    & $N_{\rm flux}\leq276$; ansatz~\eqref{eq:flux_ansatz}; $m_1,n_1\geq0$.
    & See main text.
    & See main text. \\
    \bottomrule
  \end{tabularx}
  \par\smallskip
  \begin{minipage}{\textwidth}
    \footnotesize
    LCS tests include conditions such as $|F_{\rm inst}|/|F_{\rm poly}|<$(threshold). The precise definitions and cuts differ among references.
  \end{minipage}
  \caption{Comparison of the solution counts and search conditions for the two-modulus model with earlier studies.
  }
  \label{tab:vacua_counts}
\end{table}
Our numerical SUSY count is smaller than several of the counts reported in Ref.~\cite{Chauhan:2025rdj}, while our non-SUSY count is larger than the reported count in Ref.~\cite{Krippendorf:2023idy}. 

\subsection{Moduli distributions and correlations}\label{subsec:moduli_statistics}

\begin{figure}[h]
    \centering
    \includegraphics[width=0.52\textwidth]{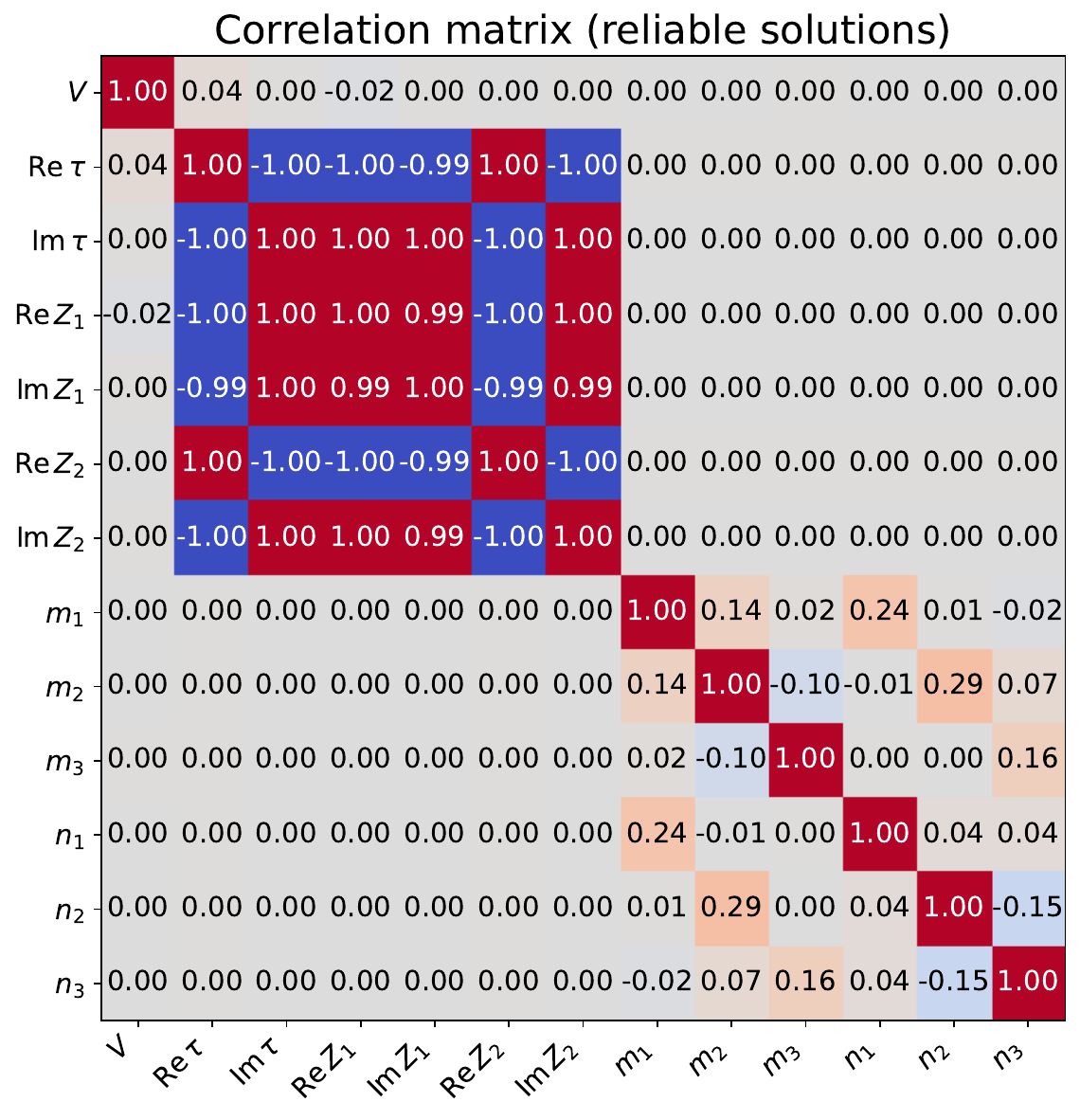}
    \includegraphics[width=0.46\textwidth]{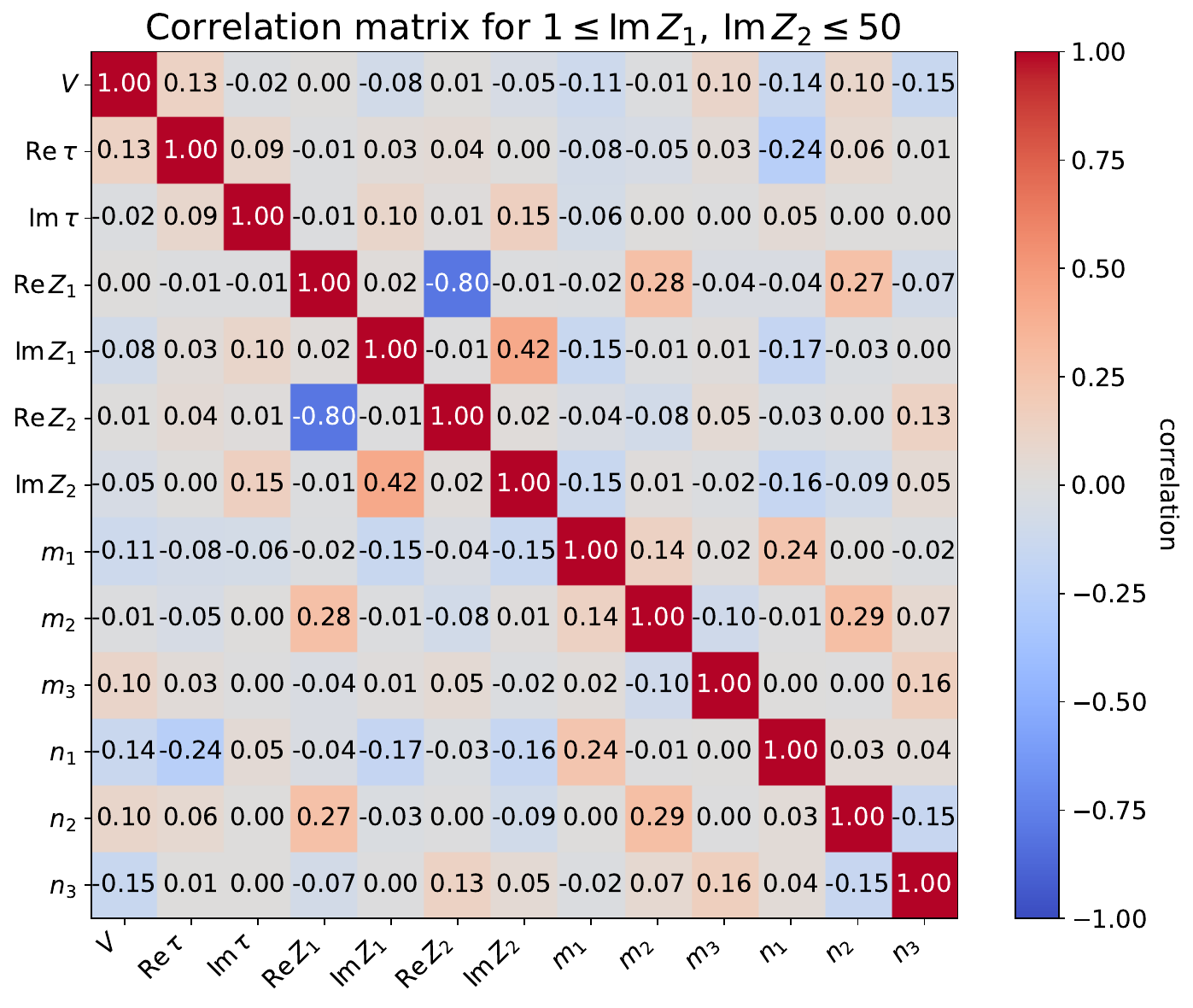}
    \caption{Pearson correlation matrices.
    Left: $1{,}296{,}633$ reliable solutions, including those with very large field values. Right: The subset of $1{,}288{,}378$ reliable solutions with $1 \leq \mathrm{Im}\, Z_1 \leq 50$ and $1 \leq \mathrm{Im}\, Z_2 \leq 50$.
    }
    \label{fig:correlation_matrix_subset}
\end{figure}
Here we study the moduli field values and their correlations.
In order to study correlations among the moduli, we compute the correlation matrices for the reliable solutions.
The Pearson correlation matrix for reliable solutions is shown in Figure~\ref{fig:correlation_matrix_subset}.
The Pearson correlation coefficients are defined as follows.
Let $X_1,\dots, X_p$ be real-valued random variables with $\mu_i$ and $\sigma_i^2$ being the means and variances, respectively.
The covariance between $X_i$ and $X_j$ is defined as
\begin{align}
\mathrm{Cov}(X_i,X_j)
= \mathbb{E}\big[(X_i-\mu_i)(X_j-\mu_j)\big] \;.
\end{align}
The correlation matrix of $(X_1,\dots,X_p)$ is then
\begin{align}
    &\rho_{ij}
    = \frac{\mathrm{Cov}(X_i,X_j)}{\sigma_i \sigma_j} \;,
    \qquad -1 \le \rho_{ij} \le 1 \;.
\end{align}
It is a positive semidefinite symmetric matrix with unit diagonal entries, i.e.\ $\rho_{ii}=1$ for all $i$.
The correlations close to $+1$ or $-1$ in the full sample become substantially weaker in the subset with $\mathrm{Im}\,Z_1,\mathrm{Im}\,Z_2\leq50$.
The real parts retain a strong anti-correlation,
$\rho_{\mathrm{Re}\,Z_1,\mathrm{Re}\,Z_2}\simeq-0.80$.

\begin{figure}[htbp]
    \centering
    \includegraphics[width=\textwidth]{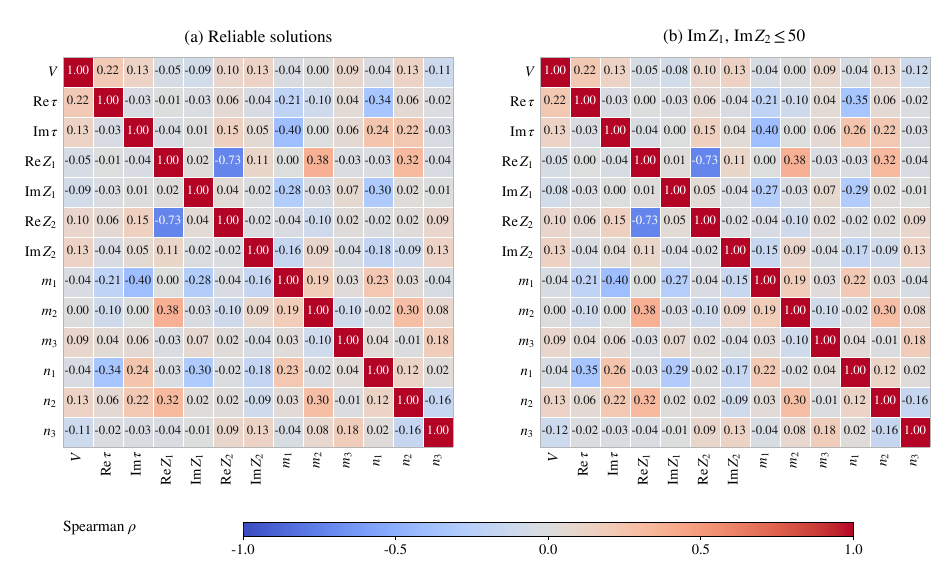}
    \caption{Spearman rank correlation matrices for the same variables, samples, and coordinates as Figure~\ref{fig:correlation_matrix_subset}.}
    \label{fig:correlation_spearman}
\end{figure}

For the assessment of the sensitivity to large field values, we also compute the Spearman rank correlations, shown in Figure~\ref{fig:correlation_spearman}.
Spearman correlation is the Pearson correlation of the ranks, with average ranks assigned to tied values.
For $\mathrm{Re}\,Z_1$ and $\mathrm{Re}\,Z_2$, the Spearman coefficients are $-0.729$ and $-0.733$ for reliable solutions and for the restricted subset, respectively.
The real-part anti-correlation therefore persists.
For $\mathrm{Im}\,Z_1$ and $\mathrm{Im}\,Z_2$, however, the Spearman coefficients are only $-0.021$ and $-0.036$, respectively.
Thus, the large Pearson coefficients for the imaginary parts are not accompanied by strong rank correlations.

\begin{figure}[h]
    \centering
    \includegraphics[width=0.49\textwidth]{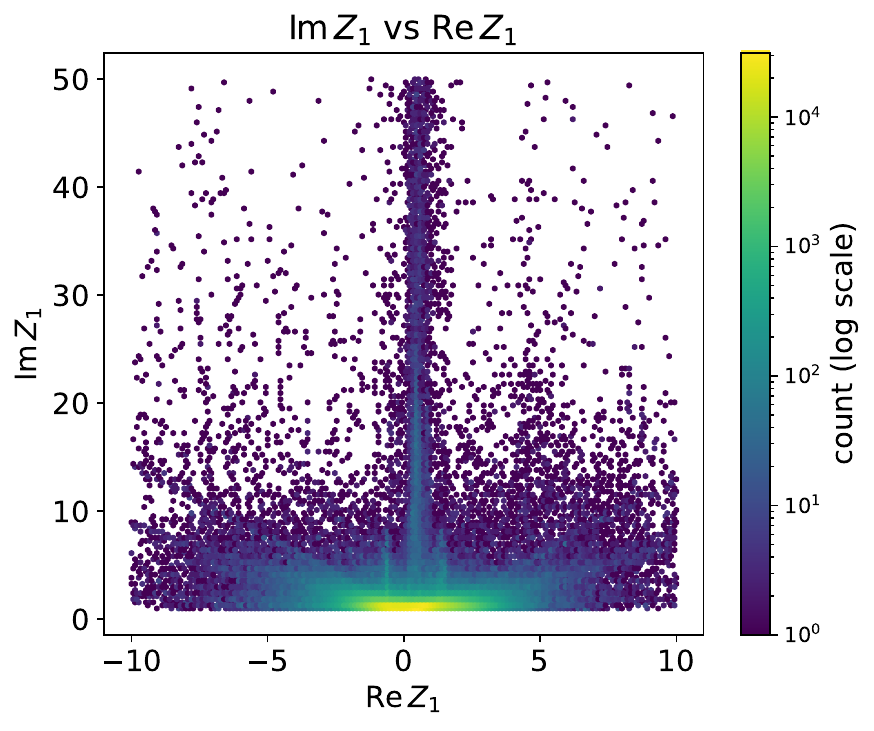}
    \includegraphics[width=0.49\textwidth]{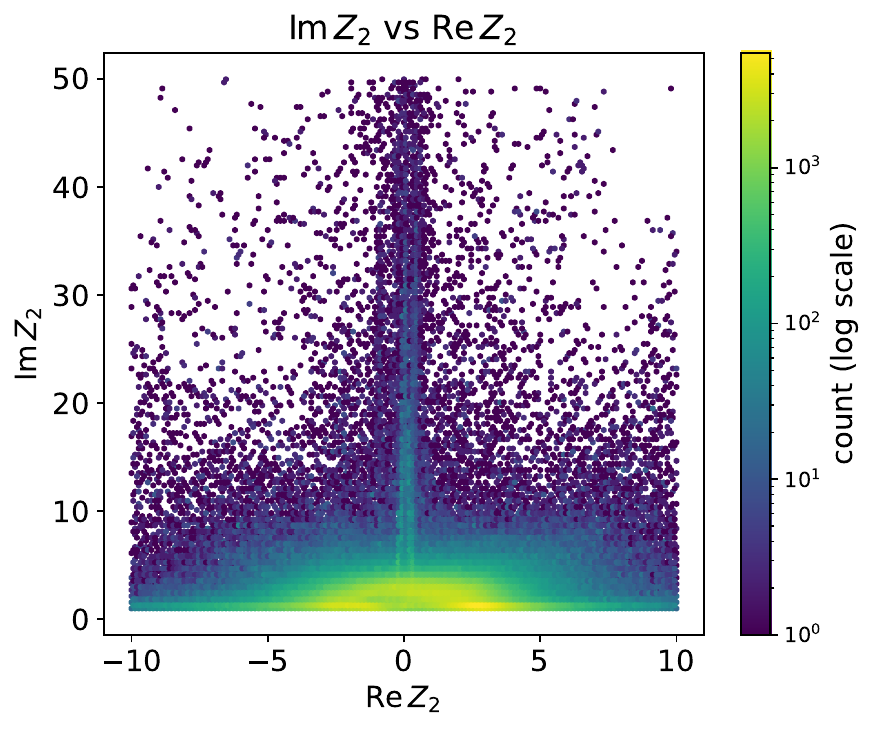}\\
    \includegraphics[width=0.49\textwidth]{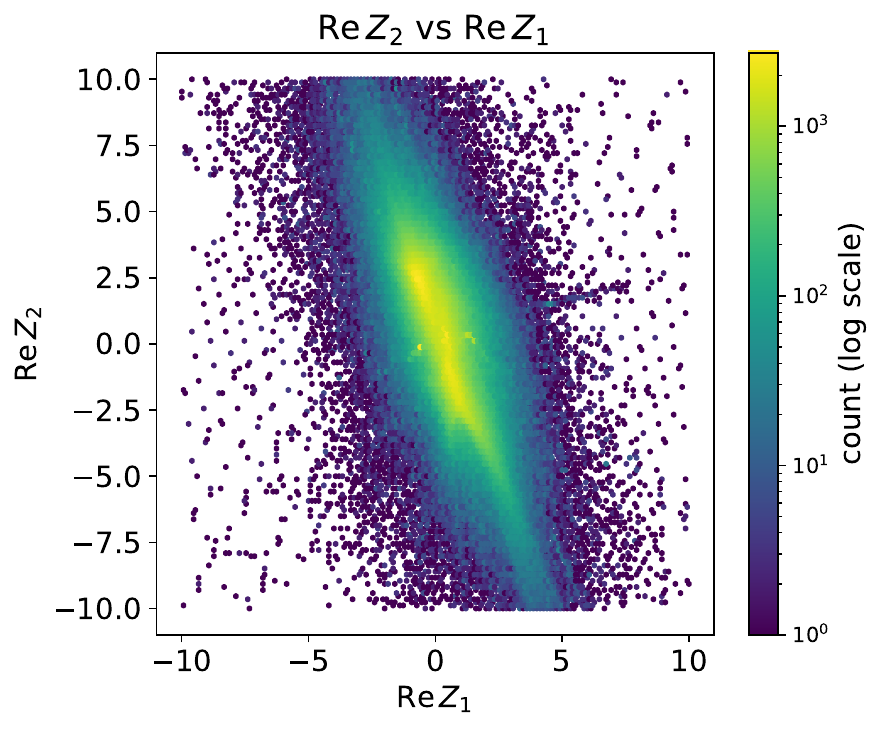}
    \includegraphics[width=0.49\textwidth]{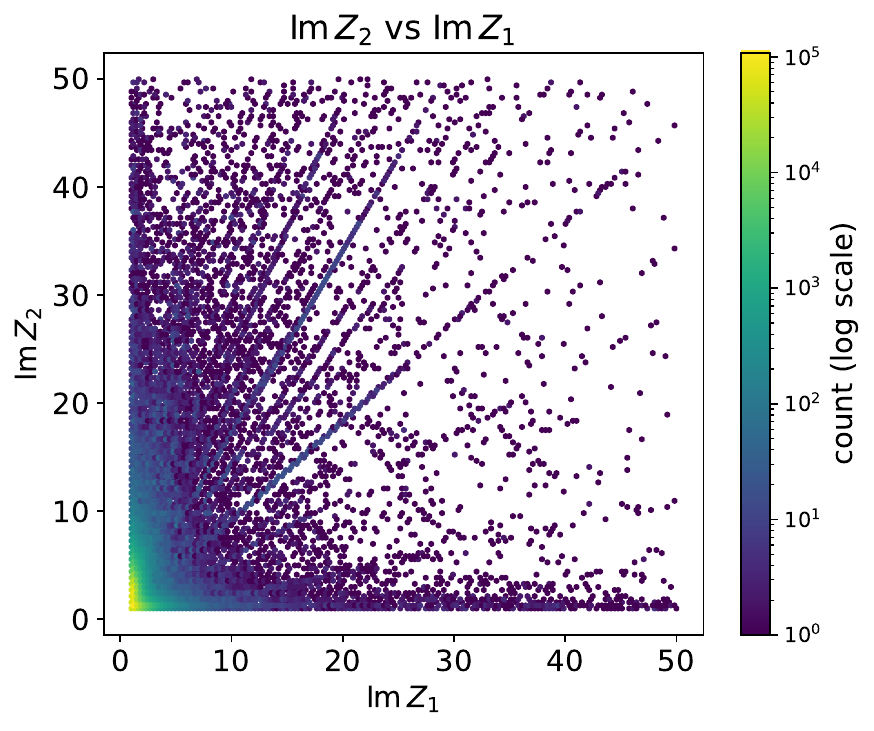}
    \caption{
Two-dimensional density plots for the reliable solutions.
We focus on the solutions with real parts in $[-10,10]$ and imaginary parts in $[1,50]$.
The fractions of solutions outside these ranges are $0.512\%$, $2.030\%$, $1.634\%$, and $0.637\%$ for the top-left, top-right, bottom-left, and bottom-right panels, respectively.
}
    \label{fig:reliable_Zi_ReIm}
\end{figure}

A correlation is more clearly visualized in two-dimensional density plots, as shown in Figure~\ref{fig:reliable_Zi_ReIm}.
We see that the real parts display a strong anti-correlation.
On the other hand, the other pairs show no clear correlation, although large imaginary parts tend to accompany small absolute values of the corresponding real parts.

\begin{table}[h]
\centering
\begin{tabular}{lrrrrrr}
\toprule
 & $\mathrm{Re}\,\tau$ & $\mathrm{Im}\,\tau$ & $\mathrm{Re}\,Z_1$ & $\mathrm{Im}\,Z_1$ & $\mathrm{Re}\,Z_2$ & $\mathrm{Im}\,Z_2$ \\
\midrule
min & -56.1 & $4.17\times10^{-3}$ & -42.1 & 1.00 & -85.9 & 1.00 \\
25\% & -1.60 & 0.188 & -0.542 & 1.09 & -1.95 & 1.16 \\
50\% & -0.732 & 0.471 & 0.225 & 1.23 & -0.419 & 1.45 \\
75\% & -0.183 & 1.41 & 0.868 & 1.54 & 2.05 & 2.22 \\
max & 86.4 & $1.72\times10^{3}$ & 41.2 & 634 & 90.5 & $1.59\times10^{3}$ \\
\bottomrule
\end{tabular}
\caption{Summary statistics for the SUSY solutions.}
\label{tab:stats_susy_fd_I}
\end{table}
\begin{figure}[h]
    \centering
    \includegraphics[width=0.7\textwidth]{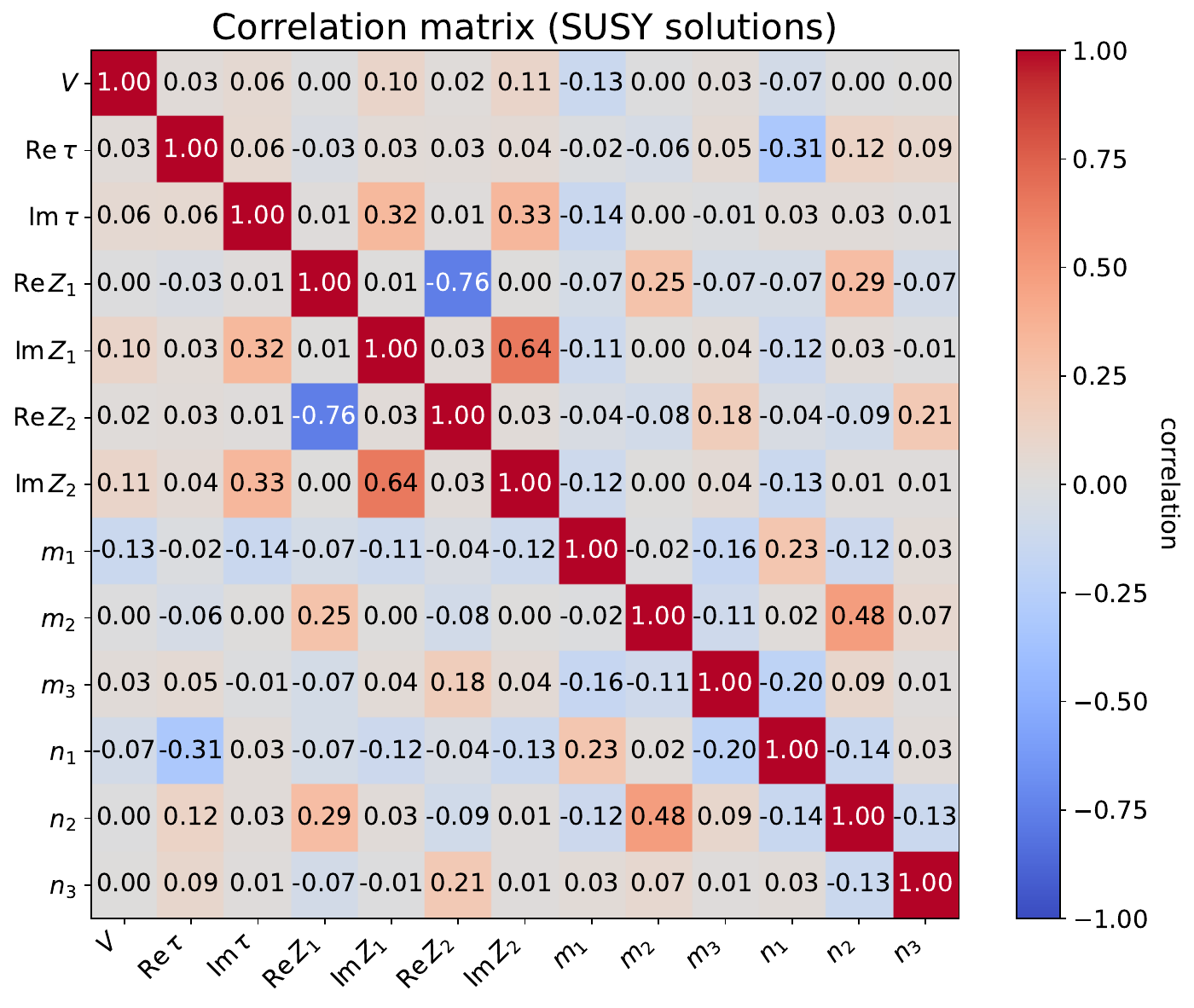}
    \caption{The Pearson correlation matrix for the $127{,}241$ SUSY solutions.}
    \label{fig:correlation_matrix_SUSY}
\end{figure}
Next, we restrict ourselves to the SUSY subset defined by Eq.~\eqref{eq:vacuum_criterion}.
Table~\ref{tab:stats_susy_fd_I} summarizes the distributions of the solutions in this subset.
The Pearson correlation matrix for the SUSY solutions is shown in Figure~\ref{fig:correlation_matrix_SUSY}.
For the real and imaginary parts of the complex-structure moduli, the Pearson coefficients are
\begin{align}
    &\rho_{\mathrm{Re}\,Z_1,\mathrm{Re}\,Z_2}=-0.76\;,\qquad
    &&\rho_{\mathrm{Im}\,Z_1,\mathrm{Im}\,Z_2}=0.64\;.
\end{align}
Figure~\ref{fig:correlation_rank_SUSY} shows the Spearman rank correlations for the SUSY sample. The real parts of $Z_1$ and $Z_2$ are anti-correlated, with a coefficient of $-0.697$, while the imaginary parts show a weaker positive correlation of $0.238$.
The Kendall rank correlations show the same qualitative trends in all three samples.

\begin{figure}[h]
    \centering
    \includegraphics[width=0.49\textwidth]{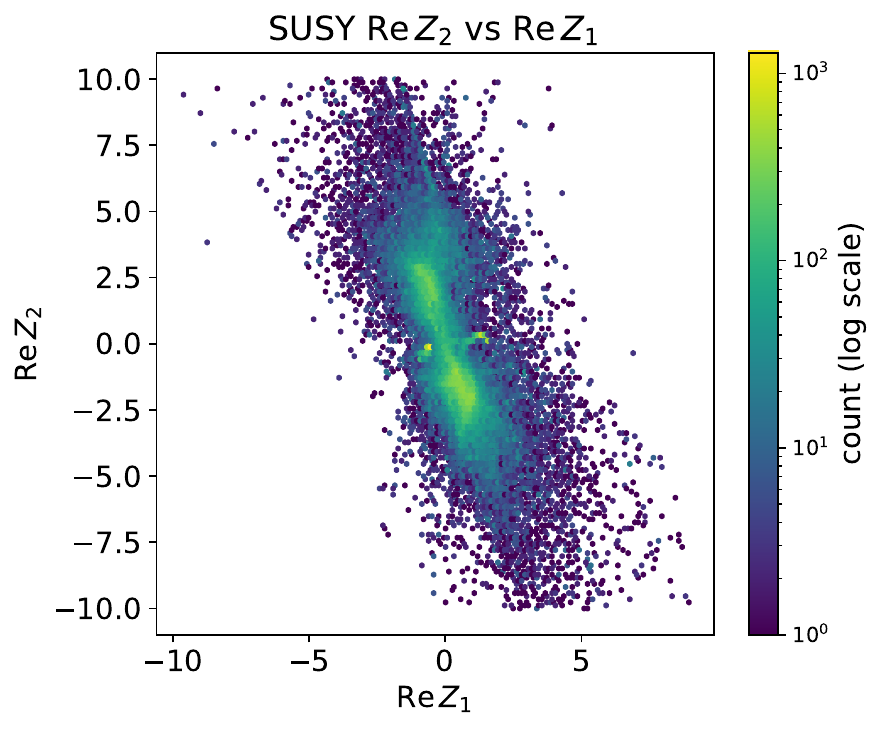}
    \includegraphics[width=0.49\textwidth]{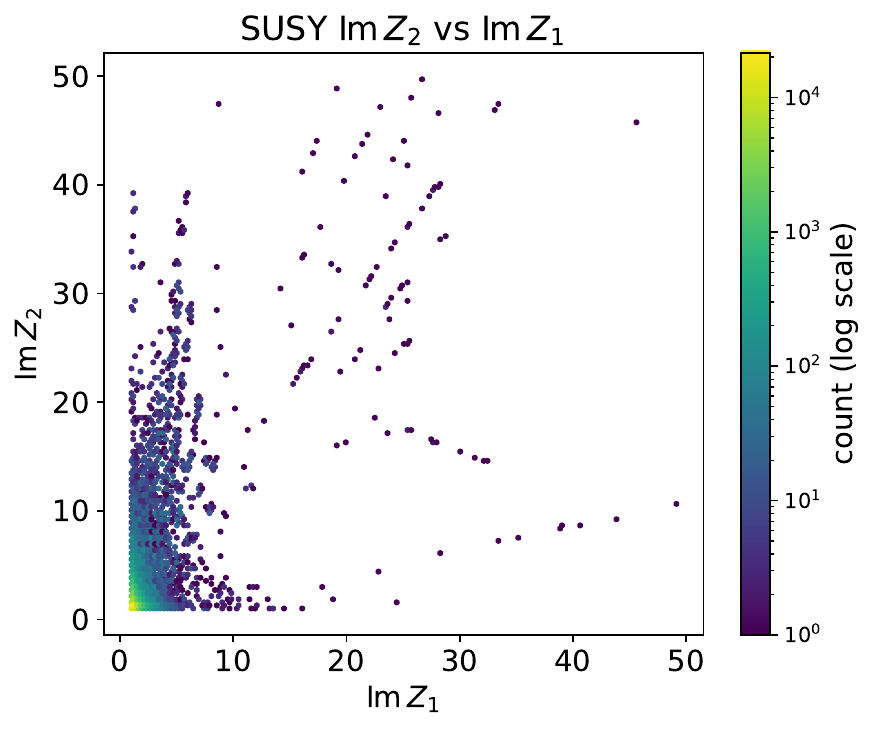}
    \caption{Two-dimensional density plots for the $127{,}241$ SUSY solutions. Left: both real parts
    are in $[-10,10]$. Right: both imaginary parts are in $[1,50]$.
    The fractions of this SUSY sample outside these ranges are $1.161\%$
    and $0.0747\%$, respectively.}
    \label{fig:SUSY_solutions_plots}
\end{figure}

Figure~\ref{fig:SUSY_solutions_plots} visualizes these relations.
The imaginary parts are concentrated near $\mathrm{Im}\,Z_i=\mathcal{O}(1)$, with a sparse tail toward larger values.
Excluding the $95$ solutions with $\max(\mathrm{Im}\,Z_1,\mathrm{Im}\,Z_2)>50$ lowers their Pearson coefficient from $0.644$ to $0.537$, while the Spearman coefficient changes only from $0.2380$ to $0.2364$, indicating that the tail primarily affects the Pearson correlation.

\begin{figure}[htbp]
    \centering
    \includegraphics[width=0.60\textwidth]{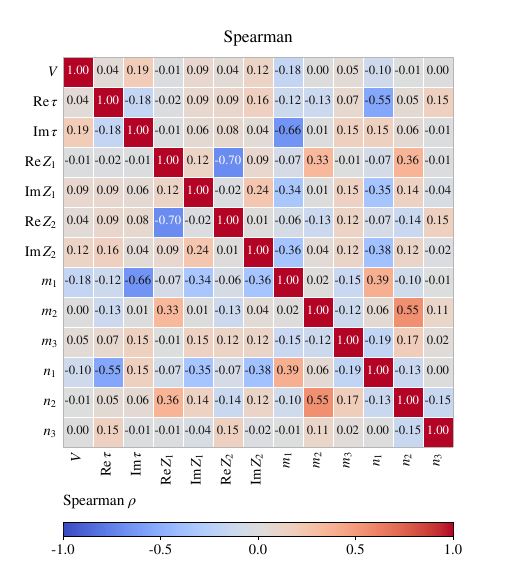}
    \caption{Spearman rank correlation matrix for all SUSY solutions.
    The variables and saved coordinates are the same as in
    Figure~\ref{fig:correlation_matrix_SUSY}; no coordinate-range cut is applied.}
    \label{fig:correlation_rank_SUSY}
\end{figure}

\subsection{Flux dependence and solution multiplicity}\label{subsec:flux_multiplicity}

We find at least one reliable solution for $1{,}147{,}948$ of the $27{,}330{,}992$ scanned $\mathbb{Z}_4$ flux orbits.
No reliable solution is found for the remaining $26{,}183{,}044$ orbits ($95.80\%$).
Since a flux choice may admit multiple solutions, we define $N_{\rm found}([\vec{N}])$ as the number of reliable solutions associated with the $\mathbb{Z}_4$ orbit $[\vec{N}]$ of $\vec{N}=(m_1,m_2,m_3,n_1,n_2,n_3)$.

\begin{table}[t]
\centering
\small
\begin{tabular}{rrr@{\qquad}rrr}
\toprule
$N_{\rm found}$ & \# flux orbits & fraction [\%] &
$N_{\rm found}$ & \# flux orbits & fraction [\%] \\
\midrule
1 & 1,063,963 & 92.6839 & 21 & 46 & 0.0040 \\
2 & 60,678 & 5.2858 & 22 & 50 & 0.0044 \\
3 & 11,968 & 1.0426 & 23 & 37 & 0.0032 \\
4 & 4,635 & 0.4038 & 24 & 23 & 0.0020 \\
5 & 2,367 & 0.2062 & 25 & 20 & 0.0017 \\
6 & 1,248 & 0.1087 & 26 & 18 & 0.0016 \\
7 & 764 & 0.0666 & 27 & 18 & 0.0016 \\
8 & 490 & 0.0427 & 28 & 12 & 0.0010 \\
9 & 327 & 0.0285 & 29 & 15 & 0.0013 \\
10 & 208 & 0.0181 & 30 & 12 & 0.0010 \\
11 & 148 & 0.0129 & 31 & 9 & 0.0008 \\
12 & 139 & 0.0121 & 32 & 7 & 0.0006 \\
13 & 112 & 0.0098 & 33 & 8 & 0.0007 \\
14 & 83 & 0.0072 & 34 & 7 & 0.0006 \\
15 & 84 & 0.0073 & 35 & 13 & 0.0011 \\
16 & 72 & 0.0063 & 36 & 6 & 0.0005 \\
17 & 81 & 0.0071 & 37 & 8 & 0.0007 \\
18 & 92 & 0.0080 & 38 & 7 & 0.0006 \\
19 & 86 & 0.0075 & 39 & 9 & 0.0008 \\
20 & 75 & 0.0065 & 40 & 3 & 0.0003 \\
\bottomrule
\end{tabular}
\caption{Distribution of $N_{\rm found}$ after duplicate removal within
each $\mathbb{Z}_4$ flux orbit. Fractions are computed among the $1{,}147{,}948$
flux orbits for which at least one reliable solution is found.}
\label{tab:nsol_per_flux}
\end{table}

The distribution of $N_{\rm found}([\vec{N}])$ is shown in Table~\ref{tab:nsol_per_flux}.
Among flux orbits for which at least one reliable solution is found, $92.68\%$ have exactly one and the rest have multiple solutions.
The largest observed value is $N_{\rm found}=40$.
An orbit can contain up to four scanned flux representatives, each with ten initial conditions, so this value is not a real upper bound on the number of vacua.

\begin{figure}[ht]
 \centering
 \includegraphics[width=\textwidth]{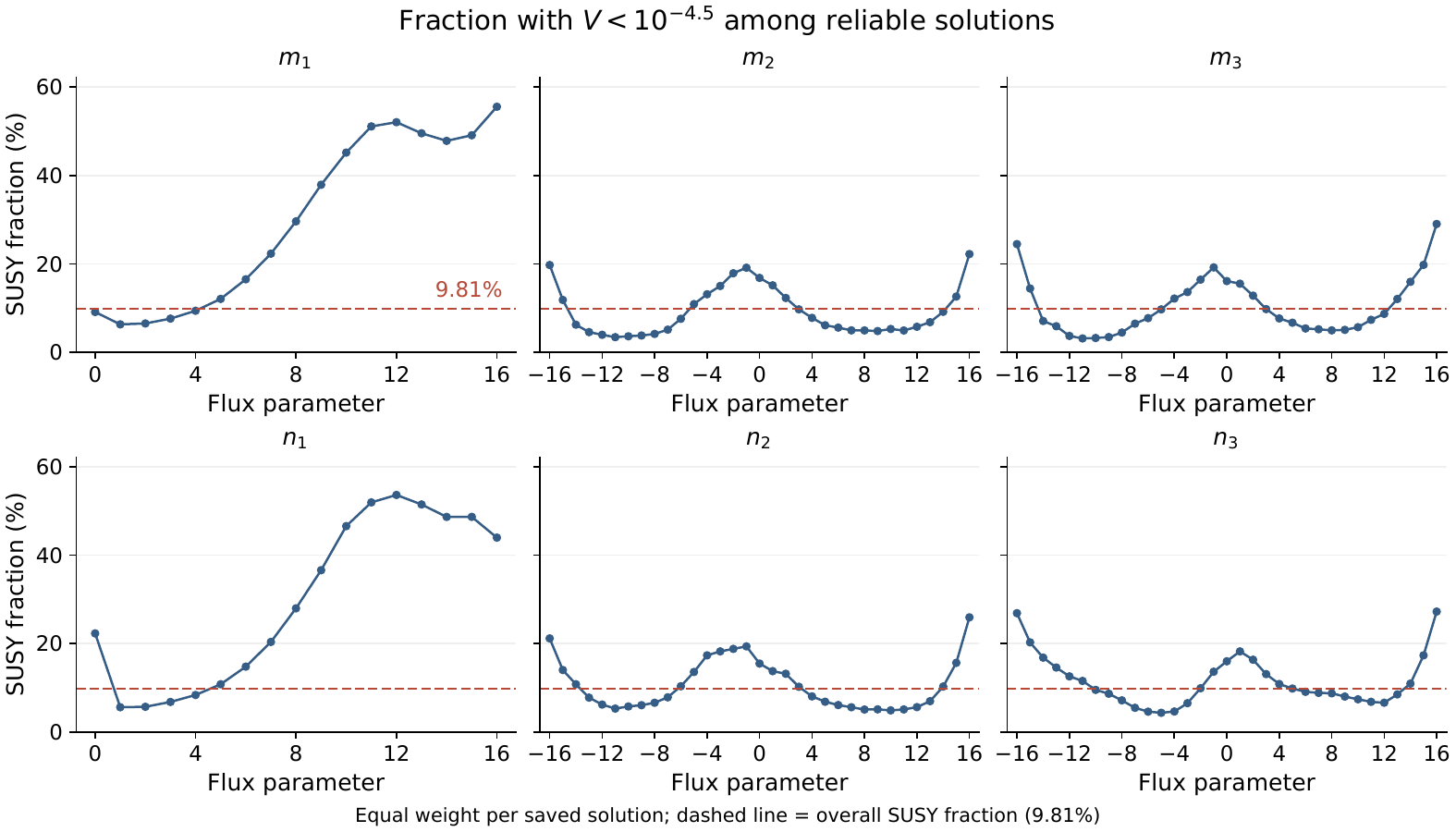}
 \caption{SUSY fraction ($V<V_*$) as a function of each flux parameter. The dashed line gives the overall SUSY fraction, 9.81\%. 
 }
 \label{fig:ml_vacuum_rate}
\end{figure}

Both SUSY and non-SUSY solutions are found for $12{,}476$ flux vectors.
For example,
\begin{align}
 \vec N=(0,-4,13,0,-2,7)\label{eq:mixed_example}
\end{align}
has solutions with $V=1.49\times10^{-6}$ and $V=1.28\times10^{-1}$.
Figure~\ref{fig:ml_vacuum_rate} shows the SUSY fraction as a function of each flux parameter.

\FloatBarrier
\section{Machine-learning analysis of flux vacua}\label{sec:ml}

In this section, we apply supervised learning to the reliable solutions computed in Section~\ref{sec:results}.\footnote{For broader applications of machine learning and data science to the string landscape, see \cite{Carifio:2017bov,Ruehle:2020jrk}.}
We perform both classification and regression tasks.
The classification task predicts whether a reliable solution is SUSY or non-SUSY, comparing the six flux parameters with invariant combinations of them.
The regression task predicts the potential and moduli field values using several input representations.

The analyses in this section use \texttt{XGBoost}~3.0.5~\cite{Chen:2016btl}.\footnote{The XGBoost is an implementation of gradient-boosted decision trees.}
For reproducibility, we record the sample indices for each data split, model configurations, software versions, and numerical summaries with the analysis scripts.

\subsection{Predicting SUSY and non-SUSY solutions}\label{subsec:ml_features}\label{subsec:ml_audit}\label{subsec:ml_classifier}

The purpose of this subsection is to classify the SUSY and non-SUSY solutions using the information of the flux as input features.
To this end, we first discuss an input representation of the flux parameters.
As we have seen in Section~\ref{sec:model}, there is a $\mathbb{Z}_4$ identification of the flux vectors, generated by the transformation $S:(m_A,n_A)\mapsto(-n_A,m_A)$, with $S^2=-I$.
Therefore, instead of using the six flux parameters $(m_A,n_A)$ as input features, it is natural to consider the following quadratic combinations invariant under $S$:
\begin{align}
 G_{AB}=m_A m_B+n_A n_B,\qquad
 \omega_{AB}=m_A n_B-m_B n_A.
\label{eq:invariants}
\end{align}
These combinations are also invariant under $SO(2)$ rotations of the pairs $(m_A,n_A)$.
There are nine invariants $G_{AB}$ and $\omega_{AB}$, but they are not algebraically independent. 

These invariants discard some information, compared with the original vanilla six flux parameters.
For example, $m=(5,0,0),\ n=(0,5,0)$ and $m'=(3,-4,0),\ n'=(4,3,0)$ have identical $G_{AB}$ and $\omega_{AB}$ but belong to different $\mathbb Z_4$ orbits. 
They are related by an $SO(2)$ rotation with $(\cos\theta,\sin\theta)=(3/5,4/5)$.
Among the $1{,}296{,}633$ solutions, there are $1{,}138{,}963$ distinct nine-invariant combinations.
Of these, $91{,}014$ contain multiple solutions, accounting for $248{,}684$ solutions in total.
More than one distinct flux vector is contained in $24{,}482$ combinations.
We will see that, despite this information loss, classifiers using the nine invariants outperform those using the six flux parameters under the model settings considered here.
Note that, because both SUSY and non-SUSY solutions can occur at the same flux vector, neither the six flux parameters nor their invariants necessarily determine whether an individual solution is SUSY.
Indeed, $12{,}166$ combinations of $G_{AB}$ and $\omega_{AB}$
contain both SUSY and non-SUSY solutions.

For classification, we use the $75\%$ of the data for training ($972{,}474$ solutions) and $25\%$ for testing ($N_{\rm test}=324{,}159$ solutions).
The test set contains $N_{{\rm SUSY,test}}=31{,}810$ SUSY solutions.
For the comparison of input representations, among the training data, we further split $75\%$ for training ($729{,}355$ solutions) and $25\%$ for validation ($243{,}119$ solutions).
Both splits preserve approximately the same SUSY fraction.

The binary classifiers use maximum depth $8$, learning rate $0.15$, minimum child weight $5$, and subsampling fractions of $0.8$ for both training solutions and input variables.
Unless otherwise stated, the classifiers use $300$ trees.

We evaluate classification performance using the following four metrics:
\begin{itemize}
    \item {\it ROC--AUC}: Measures how well the classifier ranks SUSY solutions above non-SUSY solutions (a perfect ranking gives $1$, while random ranking gives $0.5$ on average).
    \item {\it Recall}: The fraction of all SUSY solutions in the validation or test set recovered when selecting a fraction $r$ of that set in decreasing order of predicted SUSY probability.
    \item {\it Average precision}: We rank solutions by predicted SUSY probability and consider the top one, top two, and so on, including all solutions with equal scores together. At each step, precision is the fraction of these solutions that are SUSY. For each SUSY solution, we record the precision at the step when it is included. Average precision is the mean of these values. Since the SUSY fraction in the test set is $0.09813$, the random-ranking baseline is this fraction.
    \item {\it Brier score}: The mean squared difference between the predicted SUSY probability and the actual outcome, encoded as $1$ for SUSY and $0$ for non-SUSY. Predicting the same probability for every solution, equal to the SUSY fraction in the training set, gives a baseline score of $8.85\times10^{-2}$ on the test set. Smaller values indicate better predictions. We have not performed a separate calibration study.
\end{itemize}

\begin{table}[ht]
\centering
\small\setlength{\tabcolsep}{4pt}
\begin{tabular}{lrrrrr}
\toprule
input features & \# feat. & ROC--AUC & avg.\ prec. & Brier score & recall at $r=5\%$ \\
\midrule
$m_A,n_A$ & 6 & 0.9794 & 0.8244 & 0.03425 & 0.449 \\
$\omega_{AB}$, $|\vec N|^2$ & 4 & 0.9113 & 0.5799 & 0.05893 & 0.347 \\
$\omega_{AB}$, $G_{AB}$ & 9 & 0.9898 & 0.8932 & 0.02410 & 0.470 \\
$m_A,n_A,\omega_{AB}$, $|\vec N|^2$ & 10 & 0.9869 & 0.8756 & 0.02736 & 0.467 \\
$m_A,n_A,\omega_{AB}$, $G_{AB}$ & 15 & 0.9896 & 0.8953 & 0.02429 & 0.473 \\
\bottomrule
\end{tabular}
\caption{Comparison of input feature sets on the validation set.}
\label{tab:ml_features_ablation}
\end{table}

Table~\ref{tab:ml_features_ablation} compares five flux representations using the training and validation sets defined above.
We use $G_{AB},\omega_{AB}$ for the main classifier because its validation average precision, $0.8932$, is close to $0.8953$ obtained with fifteen inputs, while requiring fewer input variables.
We train this classifier on the full training set with $400$ trees and evaluate its predictions and ranking on the test set, obtaining the following scores:
\begin{align}
\text{ROC--AUC}=0.9910\;,\qquad
\text{average precision}=0.9031\;,\qquad
\text{Brier score}=2.25\times10^{-2}\;.
\label{eq:ml_metrics}
\end{align}

At selection fraction $r$, we select $\lfloor rN_{\rm test}\rfloor$ solutions from the test set in decreasing score order.\footnote{If only some solutions with the same predicted score can be selected, we average the SUSY count over all equally likely selections from that group. For example, selecting $3$ out of $10$ tied solutions containing $4$ SUSY solutions contributes $3\times4/10=1.2$ to the expected count.}
The top $5\%$ contains 47.2\% of the SUSY solutions in the test set, with precision $0.926$ (Table~\ref{tab:ml_budget}).
Even a perfect ranking obeys
\begin{align}
 \mathrm{recall}(r)\leq
 \min\!\left(1,\frac{\lfloor rN_{\rm test}\rfloor}{N_{{\rm SUSY,test}}}\right).
 \label{eq:ml_recall_bound}
\end{align}
At $r=5\%$, this upper bound is 50.9\%.
The observed recall of $47.2\%$ is therefore close to the maximum possible value at this selection fraction.

\begin{table}[ht]
\centering
\small\setlength{\tabcolsep}{4pt}
\begin{tabular}{rrrrr}
\toprule
selection fraction $r$ & solutions selected & expected SUSY solutions & recall & precision \\
\midrule
$0.1\%$ & 324 & 317 & 0.010 & 0.978 \\
$0.2\%$ & 648 & 633 & 0.020 & 0.977 \\
$0.5\%$ & 1{,}620 & 1{,}575 & 0.050 & 0.972 \\
$1\%$ & 3{,}241 & 3{,}136 & 0.099 & 0.968 \\
$2\%$ & 6{,}483 & 6{,}215 & 0.195 & 0.959 \\
$5\%$ & 16{,}207 & 15{,}012.7 & 0.472 & 0.926 \\
$10\%$ & 32{,}415 & 27{,}214 & 0.856 & 0.840 \\
$20\%$ & 64{,}831 & 31{,}769 & 0.999 & 0.490 \\
\bottomrule
\end{tabular}
\caption{Ranking of the test solutions by predicted SUSY probability.}
\label{tab:ml_budget}
\end{table}

\begin{figure}[ht]
 \centering
 \includegraphics[width=0.78\textwidth]{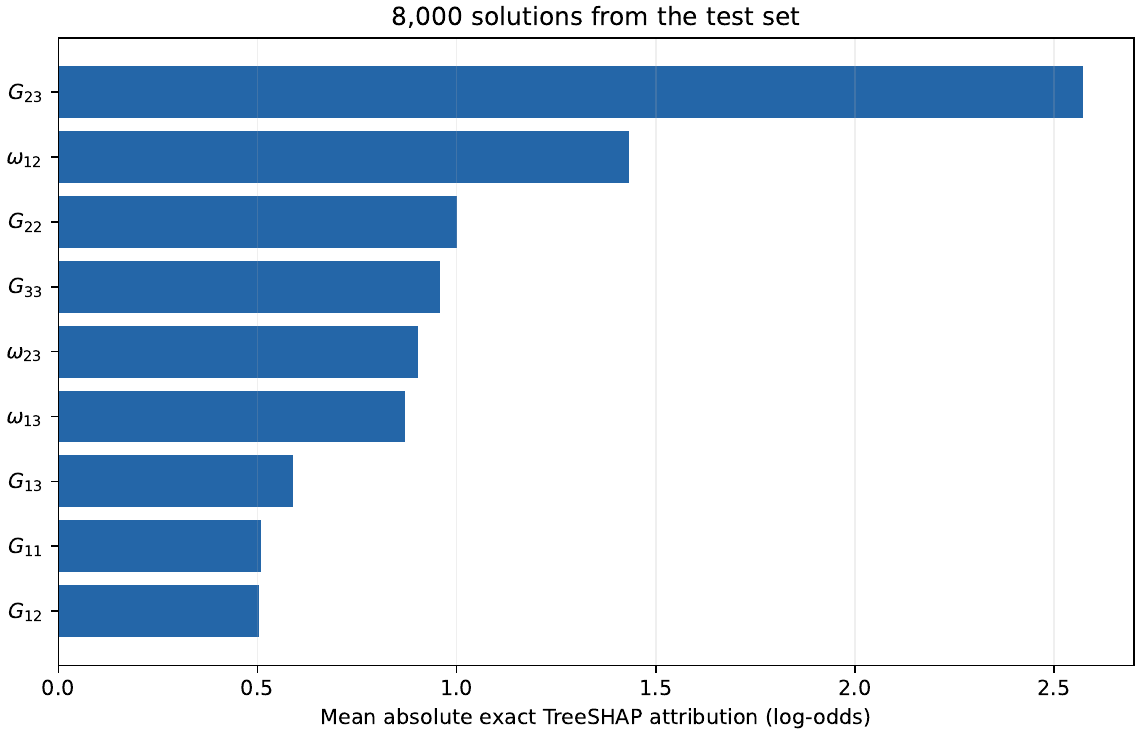}
 \caption{Mean absolute TreeSHAP contributions~\cite{Lundberg:2020tsh} defined in Eq.~\eqref{eq:shap_mean_absolute} for the classifier with
 $G_{AB},\omega_{AB}$ as input features.
 Values are averaged over $8{,}000$ test solutions and reported in log-odds units.}
 \label{fig:ml_shap}
\end{figure}

To identify which input variables contribute most to the classifier's predictions, we use TreeSHAP. For each solution $i$, TreeSHAP decomposes the predicted log-odds as
\begin{align}
\log\frac{p_i}{1-p_i}=\phi_0+\sum_{\alpha=1}^{9}\phi_{i\alpha},
\label{eq:shap_log_odds}
\end{align}
where $p_i$ is the predicted SUSY probability, $\phi_0$ is the baseline value, and $\phi_{i\alpha}$ is the contribution of input variable $\alpha$ to the prediction for solution $i$.
TreeSHAP averages over unspecified inputs using the branch weights stored in the trained trees.
With all inputs unspecified, the averaged model output gives $\phi_0$. The contribution $\phi_{i\alpha}$ is the change in predicted log-odds when input $\alpha$ is specified at its value for solution $i$, averaged over all possible orders in which the nine inputs can be specified.
The values shown in Figure~\ref{fig:ml_shap} are the mean absolute contributions over $8{,}000$ test solutions,
\begin{align}
\left\langle |\phi_\alpha|\right\rangle
=\frac{1}{8000}\sum_{i=1}^{8000}|\phi_{i\alpha}|.
\label{eq:shap_mean_absolute}
\end{align}
Larger values indicate a greater mean absolute contribution to the predicted
log-odds, compared with the other input variables in the same model.
The largest contributions are $G_{23}$ ($2.571$), $\omega_{12}$ ($1.432$),
and $G_{22}$ ($1.002$).

We next examine whether the classifier trained to identify solutions with
$V<10^{-4.5}$ is also useful for identifying solutions satisfying stricter
thresholds on $V$, without retraining (Table~\ref{tab:ml_robustness}).
For $V<10^{-8}$, selecting the top $5\%$ of the test set recovers $42.9\%$
of the $42$ solutions satisfying this threshold.

\begin{table}[ht]
\centering
\small\setlength{\tabcolsep}{4pt}
\begin{tabular}{lrrrrr}
\toprule
criterion & \shortstack{test solutions\\below threshold} & \shortstack{fraction below\\threshold} & ROC--AUC & avg.\ prec. & recall at $r=5\%$ \\
\midrule
$V<10^{-4.5}$ & 31{,}810 & $9.81\times10^{-2}$ & 0.9910 & 0.9031 & 0.472 \\
$V<10^{-5}$ & 29{,}759 & $9.18\times10^{-2}$ & 0.9885 & 0.8655 & 0.483 \\
$V<10^{-6}$ & 10{,}305 & $3.18\times10^{-2}$ & 0.9630 & 0.3627 & 0.550 \\
$V<10^{-7}$ & 1{,}029 & $3.17\times10^{-3}$ & 0.9515 & 0.0439 & 0.587 \\
$V<10^{-8}$ & 42 & $1.30\times10^{-4}$ & 0.9420 & 0.0029 & 0.429 \\
\bottomrule
\end{tabular}
\caption{We evaluate the classifier using $G_{AB},\omega_{AB}$ as inputs on the same test set at stricter thresholds on $V$, without retraining.}
\label{tab:ml_robustness}
\end{table}

Because multiple reliable solutions can have the same flux vector, identical input feature values can appear in both the training and test sets when individual solutions are randomly assigned to these sets.
In this split ({\it Individual solution} in Table~\ref{tab:ml_grouped}), 16.39\% of the test solutions have the same values of $\omega_{AB},G_{AB}$ as at least one training solution. This fraction is 57.40\% among SUSY solutions in the test set.
We therefore repeat the comparison between $m_A,n_A$ and $\omega_{AB},G_{AB}$ by keeping all solutions with the same invariant values together in either the training or test set ({\it Invariant group} in Table~\ref{tab:ml_grouped}).
We assign $75\%$ of these groups to training and $25\%$ to testing, preserving approximately the same fraction of groups containing at least one SUSY solution in both sets.
For comparison, we then select samples with the same numbers of training and test solutions under the two ways of splitting the data, while approximately preserving the SUSY fraction of each original set.
For each way, both input representations use the same training and test solutions.

\begin{table}[ht]
\centering
\small\setlength{\tabcolsep}{1.8pt}
\begin{tabular}{llrrrr}
\toprule
Split method & input & SUSY fraction & ROC--AUC & avg.\ prec. & recall at $r=5\%$ \\
\midrule
Individual solution & $m_A,n_A$ & $9.813\%$ & 0.9799 & 0.8286 & 0.450 \\
Individual solution & $\omega_{AB},G_{AB}$ & $9.813\%$ & 0.9904 & 0.8986 & 0.472 \\
Invariant group & $m_A,n_A$ & $9.758\%$ & 0.9780 & 0.8170 & 0.448 \\
Invariant group & $\omega_{AB},G_{AB}$ & $9.758\%$ & 0.9882 & 0.8839 & 0.473 \\
\bottomrule
\end{tabular}
\caption{Comparison of two split methods. An invariant group consists of all solutions with identical values of the nine invariants $G_{AB},\omega_{AB}$.
Under the split by invariant group, the same $(\omega_{AB},G_{AB})$ values appear in only one of the training or test sets.}
\label{tab:ml_grouped}
\end{table}

The result of the comparison is shown in Table~\ref{tab:ml_grouped}.
Under the split by invariant group, average precision is lower for both input representations, while recall at $r=5\%$ remains almost unchanged.
The classifier using $G_{AB},\omega_{AB}$ achieves higher average precision under both split methods.
We also observe that the split by invariant group does not increase the performance of the classifier.

Finally, returning to the original split by individual solution, we examine how prediction performance changes as the training set grows, using $G_{AB},\omega_{AB}$ as input features and evaluating all models on the test set from that split.
Average precision increases from $0.7278$ to $0.8974$ as the training set grows from $10{,}000$ to $972{,}474$ solutions (Table~\ref{tab:ml_learning_curve}).

\begin{table}[ht]
\centering
\small\setlength{\tabcolsep}{4pt}
\begin{tabular}{lrrr}
\toprule
training solutions & ROC--AUC & avg.\ prec. & recall at $r=5\%$ \\
\midrule
10{,}000 & 0.9661 & 0.7278 & 0.397 \\
40{,}000 & 0.9786 & 0.8133 & 0.439 \\
160{,}000 & 0.9857 & 0.8637 & 0.459 \\
972{,}474 & 0.9902 & 0.8974 & 0.471 \\
\bottomrule
\end{tabular}
\caption{Prediction performance as the number of training solutions increases.
Training sets are sampled from the main training set. Each larger training set contains all solutions from the smaller training sets, and all training sets have approximately the same SUSY fraction. Every model is evaluated on the same test set.}
\label{tab:ml_learning_curve}
\end{table}

\subsection{Predicting the potential and moduli}\label{subsec:ml_regression}

We consider two regression tasks using $1{,}296{,}633$ reliable solutions:
\begin{itemize}
\item {\it Potential prediction}: We predict $\log_{10}V$, comparing four sets of input features.
\item {\it Moduli prediction}: We predict the six moduli coordinates using either $m_A,n_A$ or $G_{AB},\omega_{AB}$ as input features.
\end{itemize}
For potential prediction, we use $300$ trees with learning rate $0.15$.
For moduli prediction, we use $400$ trees with learning rate $0.1$.
Both tasks use maximum depth $8$, minimum child weight $5$, and subsampling fractions of $0.8$ for both training solutions and input features. We randomly select $75\%$ of the solutions for training and use the remaining $25\%$ for testing. All models in both tasks use the same training and test sets.
As in classification, the training and test sets can contain solutions with
identical values of the flux input features.
The fraction of solutions in the test set sharing an exact flux vector with the training set is $13.77\%$ for both tasks.

In order to evaluate regression performance, we use two metrics.
One is the coefficient of determination
\begin{align}
R^2=1-\frac{\sum_i(y_i-\hat y_i)^2}{\sum_i(y_i-\bar y)^2},\qquad \bar y=\frac{1}{n}\sum_i y_i,\label{eq:R2}
\end{align}
and the other is the mean absolute error (MAE),
\begin{align}
\mathrm{MAE}=\frac{1}{n}\sum_{i=1}^{n}\left|y_i-\hat y_i\right|.
\label{eq:MAE}
\end{align}
Both metrics are evaluated on the test set, where $n$ is the number of test
solutions and $y_i$ and $\hat y_i$ are the target and predicted values, respectively.
Higher $R^2$ indicates better predictions: $R^2=1$ corresponds to perfect
predictions, $R^2=0$ to the same squared error as predicting the test-set mean
$\bar y$ for every solution, and $R^2<0$ to a larger squared error than this baseline.
Lower MAE is better, with $\mathrm{MAE}=0$ for perfect predictions.

When moduli are included as inputs for potential prediction, we supply the
six coordinates both directly and after the transformation
$x\mapsto\mathrm{sign}(x)\ln(1+|x|)$, giving twelve additional input features.
The logarithmic transformation compresses the wide numerical range while
preserving the signs. Table~\ref{tab:ml_logV} compares the four input
representations under identical settings, with ``$+$ moduli'' denoting
these twelve additional input features.
Adding the moduli to $G_{AB},\omega_{AB}$ increases $R^2$ from $0.886$ to
$0.909$ and reduces MAE from $0.380$ to $0.345$.

\begin{table}[ht]\centering
\begin{tabular}{lrrr}\toprule
input & \# feat. & $R^2$ & MAE in $y=\log_{10}V$ \\\midrule
$G_{AB},\omega_{AB}$ & 9 & $0.886$ & $0.380$ \\
$G_{AB},\omega_{AB}$ $+$ moduli & 21 & $0.909$ & $0.345$ \\
$m_A,n_A$ $+$ moduli & 18 & $0.844$ & $0.483$ \\
$m_A,n_A$ $+$ $G_{AB},\omega_{AB}$ $+$ moduli & 27 & $0.908$ & $0.346$ \\
\bottomrule\end{tabular}
\caption{Regression of $\log_{10}V$ on the $1{,}296{,}633$ reliable solutions.}
\label{tab:ml_logV}\end{table}

\begin{figure}[ht]\centering
\includegraphics[width=0.75\textwidth]{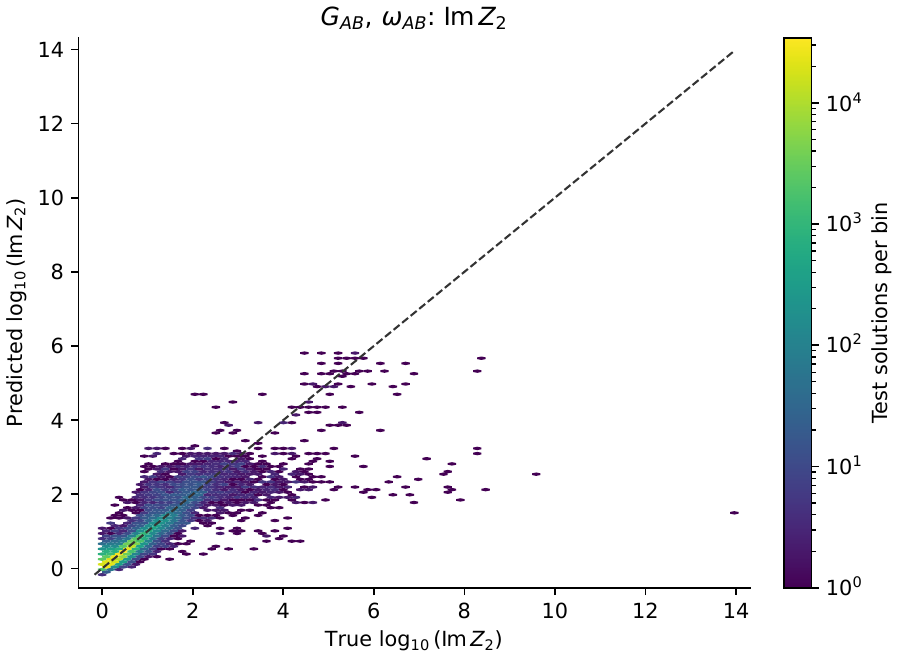}
\caption{Predicted against true values of $\log_{10}(\mathrm{Im}\,Z_2)$, the target with the largest $R^2$ for the model with $G_{AB},\omega_{AB}$ as input features. The dashed line is the identity.}
\label{fig:ml_moduli_prediction}\end{figure}

For moduli prediction, we train a separate regressor for each coordinate,
using $y=\log_{10}x$ for imaginary parts and $y=\mathrm{sign}(x)\ln(1+|x|)$ for real parts.
The $R^2$ and MAE scores are shown in Table~\ref{tab:ml_moduli}.
Using $G_{AB},\omega_{AB}$ instead of $m_A,n_A$ improves the test-set $R^2$ for all four complex-structure coordinates but reduces it for both axio-dilaton coordinates.
Among the models using $G_{AB},\omega_{AB}$ as input features, the largest $R^2$ is $0.855$ for $\log_{10}(\mathrm{Im}\,Z_2)$.
In Figure~\ref{fig:ml_moduli_prediction}, we show predicted against true values of $\log_{10}(\mathrm{Im}\,Z_2)$.

\begin{table}[ht]\centering\small\setlength{\tabcolsep}{4pt}
\begin{tabular}{lrrrrrr}\toprule
target & $\mathrm{Re}\,\tau$ & $\mathrm{Im}\,\tau$ & $\mathrm{Re}\,Z_1$ & $\mathrm{Im}\,Z_1$ & $\mathrm{Re}\,Z_2$ & $\mathrm{Im}\,Z_2$\\\midrule
$R^2$ from $m_A,n_A$ & $0.785$ & $0.910$ & $0.669$ & $0.712$ & $0.750$ & $0.763$\\
$R^2$ from $G_{AB},\omega_{AB}$ & $0.587$ & $0.692$ & $0.805$ & $0.795$ & $0.851$ & $0.855$\\
MAE from $m_A,n_A$ & $0.125$ & $0.088$ & $0.240$ & $0.060$ & $0.346$ & $0.082$\\
MAE from $G_{AB},\omega_{AB}$ & $0.294$ & $0.195$ & $0.147$ & $0.040$ & $0.215$ & $0.051$\\
\bottomrule\end{tabular}
\caption{Predicting the six coordinates~\eqref{eq:gradient_selection} for 1,296,633 reliable solutions. $R^2$ and MAE refer to transformed coordinates.}
\label{tab:ml_moduli}\end{table}

\FloatBarrier
\section{Discussion and conclusions}\label{sec:conclusion}

We have studied the landscape of flux vacua in an effective two-modulus model of type IIB orientifold compactifications.
We have generated a large dataset of $1{,}296{,}633$ reliable solutions, with $127{,}241$ SUSY solutions.
These solutions pass the selection criteria such as the tadpole bound, the smallness of the gradient, and the Hessian criterion in Eq.~\eqref{eq:hessian_cut}.
We have performed a detailed analysis of their properties.

Several features of the landscape are extracted from the dataset.
The small fraction of the SUSY solutions implies that finding them by random search is inefficient, and most of the solution space may be occupied by non-SUSY solutions.
We have also found the correlations among the moduli field values, which may be useful for future studies of flux vacua.
Moreover, even after fixing the fluxes, multiple solutions can exist.
Indeed, we have found $40$ solutions with the flux vector in the same $\mathbb Z_4$ orbit.

The machine-learning analysis has been performed to the dataset we generated.
Both classification and regression tasks have been considered.
For classification, we have trained a binary classifier to identify SUSY solutions from information of fluxes.
It has been found that, instead of using the vanilla six flux parameters, using the nine invariants under the $S$ transformation improves the classification performance.
The best classifier reaches ROC--AUC $0.9910$ and average precision $0.9031$ on the test set.
Its top $5\%$ contains 47.2\% of SUSY test solutions.

For regression, we have trained models to predict the potential and moduli field values from the information of fluxes.
Using the nine invariants as input features improves the prediction of potential values and complex-structure coordinates, while it worsens the prediction of axio-dilaton coordinates.
The best model for the potential prediction reaches $R^2=0.909$ and MAE $0.345$ on the test set.
For moduli prediction using nine-invariants as input features, the largest $R^2$ is $0.855$, with MAE $0.051$, for $\log_{10}(\mathrm{Im}\,Z_2)$.

Since the moduli field values are predicted from the fluxes with high accuracy, using the predicted moduli as initial conditions may improve the efficiency of the minimizations of the potential.
It is interesting to explore the performance such as the success rate, the number of iterations, and the running time of this approach in future work.

\acknowledgments
This work is supported by JST BOOST Program Japan Grant No.\ JPMJBY25E1 [YH], JSPS KAKENHI Grant (Nos.~JP24K07035 [YH], JP24KF0167 [YH], JP23K17689 [MY], and JP23K25865 [MY]),
JST PRESTO Grant (No.\ JPMJPR225A [MY]), and JST Moonshot R\& D Grant (No.\ JPMJMS2061 [MY]).

\section*{Declaration of AI use}
The authors used Codex (OpenAI) and Claude Code (Anthropic) to assist with developing the code for the machine-learning analyses presented in Section~\ref{sec:ml} and with polishing the language of the manuscript. The authors take full responsibility for the content of this work.

\FloatBarrier

\bibliographystyle{JHEP}
\bibliography{reference}

\end{document}

%% file: W.tex
{\small\begin{align}
&W=-\frac{1}{2} m_1 (3 Z_1^2+3 Z_1 Z_2+Z_2^2)Z_1 \tau \nonumber\\
 &+ \frac{1}{2} \left[-n_1(3 Z_1^2+3 Z_1 Z_2+Z_2^2)Z_1+(m_2 (3 Z_1+Z_2)^2 +m_3(3 Z_1+2 Z_2)Z_1) \tau\right]\nonumber\\
 &+\frac{1}{4} \left[-(17 m_1 +6 (3 m_2+m_3) -4n_2 )Z_1 \tau+(2n_3-3 (m_1+m_2) )2Z_2 \tau\right.\nonumber\\
 &\qquad\left.+2 n_2 (3 Z_1+Z_2)^2+2n_3 (3 Z_1^2+2 Z_1 Z_2)\right]\nonumber\\
 &-\frac{1}{4}\left[(4 m_2+17 n_1+18 n_2+6 n_3) Z_1+2(2m_3 +3 n_1+3 n_2)Z_2\right.\nonumber\\
 &\qquad\left.+\left(17 m_2+6 m_3-4 n_1-\frac{270i\zeta(3)}{\pi^3}m_1\right)\tau\right]\nonumber\\
 &-\frac{1}{4} \left[4 m_1+17 n_2+6 n_3-\frac{270i\zeta(3)}{\pi^3}n_1\right],\\
&\partial_\tau W=-\frac{1}{2} m_1 Z_1 (3 Z_1^2+3 Z_1 Z_2+Z_2^2)+\frac{1}{2} \left(m_2 (3 Z_1+Z_2)^2+m_3 Z_1 (3 Z_1+2 Z_2)\right)\nonumber\\
 &+\frac{1}{4} (-17 m_1-18 m_2-6m_3+4 n_2) Z_1+\frac{1}{2}(-3 m_1-3m_2+2 n_3) Z_2\nonumber\\
 &+\frac{1}{4} \left(-17 m_2-6 m_3+4 n_1+\frac{270i\zeta(3)}{\pi^3}m_1\right),\\
&\partial_{Z_1} W=-\frac{1}{2} m_1 (3 Z_1+Z_2)^2 \tau-\frac{1}{2} (3 Z_1+Z_2)(n_1(3 Z_1+Z_2) -2(3 m_2+m_3)  \tau)\nonumber\\
 &+(3 n_2+n_3) (3 Z_1+Z_2)-\frac{1}{4} (17 m_1+18 m_2+6m_3-4 n_2) \tau-\frac{1}{4} (4 m_2+17 n_1+18  n_2+6n_3),\\
&\partial_{Z_2} W=-\frac{1}{2} m_1 Z_1 (3 Z_1+2 Z_2) \tau-\frac{1}{2} n_1 Z_1 (3 Z_1+2 Z_2)+(m_3 Z_1 +m_2 (3 Z_1+Z_2)) \tau\nonumber\\
 &+ (3n_2+n_3) Z_1+n_2 Z_2-\frac{1}{2} (3m_1+3m_2-2 n_3) \tau-m_3-\frac{3}{2}(n_1+n_2),
\end{align}}

%% file: dK.tex
\begin{align}
&\partial_\tau K= -\frac{1}{\tau-\overline{\tau}},\\
&\partial_{Z_1}K = -\frac{(3 (Z_1-\overline{Z}_1)+Z_2-\overline{Z}_2)^2}{(Z_1-\overline{Z}_1) (3 (Z_1-\overline{Z}_1)^2+(Z_2-\overline{Z}_2)^2 +3(Z_1-\overline{Z}_1) (Z_2-\overline{Z}_2))-\frac{270i\zeta(3)}{\pi^3}},\\
&\partial_{Z_2}K = -\frac{(Z_1-\overline{Z}_1) (3 (Z_1-\overline{Z}_1)+2 (Z_2-\overline{Z}_2))}{(Z_1-\overline{Z}_1) (3 (Z_1-\overline{Z}_1)^2+(Z_2-\overline{Z}_2)^2+3(Z_1-\overline{Z}_1) (Z_2-\overline{Z}_2))-\frac{270i\zeta(3)}{\pi^3}},
\end{align}

%% file: K.tex
\begin{align}
&e^K = \frac{2}{(\tau-\overline{\tau}) \left((Z_1-\overline{Z}_1) (3 (Z_1-\overline{Z}_1)^2+(Z_2-\overline{Z}_2)^2+3 (Z_1-\overline{Z}_1) (Z_2-\overline{Z}_2))-\frac{270 i \zeta(3)}{\pi^3}\right)},\\
&K^{\tau\overline{\tau}}=-(\tau-\overline{\tau})^2,\\
&K^{Z_1\overline{Z}_1}=-\frac{1}{2}\frac{Z_1-\overline{Z}_1}{Z_2-\overline{Z}_2}\nonumber\\
 &\qquad\times\frac{(Z_1-\overline{Z}_1) (3 (Z_1-\overline{Z}_1)^2+(Z_2-\overline{Z}_2)^2+3(Z_1-\overline{Z}_1) (Z_2-\overline{Z}_2))-\frac{270 i \zeta(3)}{\pi^3}}{3(Z_1-\overline{Z}_1)+Z_2-\overline{Z}_2}\nonumber\\
 &\times\frac{(Z_1-\overline{Z}_1)(3 (Z_1-\overline{Z}_1)^2+2 (Z_2-\overline{Z}_2)^2+6(Z_1-\overline{Z}_1) (Z_2-\overline{Z}_2))+\frac{540 i \zeta(3)}{\pi^3}}{(Z_1-\overline{Z}_1) (3 (Z_1-\overline{Z}_1)^2+(Z_2-\overline{Z}_2)^2+3(Z_1-\overline{Z}_1) (Z_2-\overline{Z}_2))+\frac{540 i \zeta(3)}{\pi^3}},\\
&K^{Z_1\overline{Z}_2}=K^{\overline{Z}_1 Z_2}=\frac{3}{2}\frac{(Z_1-\overline{Z}_1)^2 (Z_1-\overline{Z}_1+Z_2-\overline{Z}_2)+\frac{180 i \zeta(3)}{\pi^3}}{Z_2-\overline{Z}_2} \nonumber\\
 &\times \frac{(Z_1-\overline{Z}_1) (3 (Z_1-\overline{Z}_1)^2+(Z_2-\overline{Z}_2)^2+3 (Z_1-\overline{Z}_1) (Z_2-\overline{Z}_2))-\frac{270 i \zeta(3)}{\pi^3}}{(Z_1-\overline{Z}_1) (3 (Z_1-\overline{Z}_1)^2+(Z_2-\overline{Z}_2)^2+3(Z_1-\overline{Z}_1) (Z_2-\overline{Z}_2))+\frac{540 i \zeta(3)}{\pi^3}},\\
&K^{Z_2\overline{Z}_2}=-\frac{\begin{aligned}
 &(9(Z_1-\overline{Z}_1)^3+(Z_2-\overline{Z}_2)^3\\
 &\quad+9 (Z_1-\overline{Z}_1)^2 (Z_2-\overline{Z}_2)+3(Z_1-\overline{Z}_1)(Z_2-\overline{Z}_2)^2)+\frac{1620 i \zeta(3)}{\pi^3}
 \end{aligned}}{2 (Z_2-\overline{Z}_2)}\nonumber\\
 &\times\frac{(Z_1-\overline{Z}_1) (3 (Z_1-\overline{Z}_1)^2+(Z_2-\overline{Z}_2)^2+3(Z_1-\overline{Z}_1) (Z_2-\overline{Z}_2))-\frac{270 i \zeta(3)}{\pi^3}}{(Z_1-\overline{Z}_1) (3 (Z_1-\overline{Z}_1)^2+(Z_2-\overline{Z}_2)^2+3(Z_1-\overline{Z}_1) (Z_2-\overline{Z}_2))+\frac{540 i \zeta(3)}{\pi^3}},
\end{align}